\documentclass{emulateapj}
\usepackage{color}
\usepackage{graphicx}
\newcommand{\ha}{H$\alpha$}

\newcommand{\lir}{L$\rm_{IR}$}

\newcommand{\smy}{$\rm M_\odot ~ yr^{-1}$}

\newcommand{\edi}{EDisCS}
\newcommand{\rtwo}{$\rm R_{200}$}

\newcommand{\mvcut}{$\rm M_V < -20.1$}

\shorttitle{SF in Intermediate-z Clusters}
\shortauthors{Finn et al.}

\begin{document}
\title{Dust-Obscured Star-Formation in Intermediate Redshift Galaxy Clusters}

\author{Rose A. Finn\altaffilmark{1}, 
Vandana Desai\altaffilmark{2,3}, 
Gregory Rudnick\altaffilmark{4,5}, 
Bianca Poggianti\altaffilmark{6}, 
Eric F. Bell\altaffilmark{7,8} 
Joannah Hinz\altaffilmark{9}, 
Pascale Jablonka\altaffilmark{10}, 
Bo Milvang-Jensen\altaffilmark{11}, 
John Moustakas\altaffilmark{12}, 
Kenneth Rines\altaffilmark{13},
Dennis Zaritsky\altaffilmark{9}}

\email{rfinn@siena.edu}
\altaffiltext{1}{Department of Physics, Siena College, 515 Loudon Rd., Loudonville, NY  12211}
\altaffiltext{2}{Division of Physics, Mathematics and Astronomy, California Institute of Technology, Pasadena, CA 91125}
\altaffiltext{3}{Spitzer Science Center, California Institute of Technology, Pasadena, CA 91125}
\altaffiltext{4}{University of Kansas, Department of Physics and Astronomy, Malott Hall, Room 1082, 1251 Wescoe Hall Dr., Lawrence, KS 66045}
\altaffiltext{5}{National Optical Astronomy Observatory, Tucson, AZ}
\altaffiltext{6}{Osservatorio Astronomico, vicolo dell'Osservatorio 5, 35122 Padova}
\altaffiltext{7}{Max-Planck-Institut f\"ur Astronomie, K\"onigstuhl 17, D69117 Heidelberg, Germany}
\altaffiltext{8}{Department of Astronomy, University of Michigan, 500 Church St., Ann
Arbor MI 48105}
\altaffiltext{9}{Steward Observatory, 933 N. Cherry Ave., University of Arizona, Tucson, AZ  85721}
\altaffiltext{10}{Observatoire de l’Universit\'e de Gen\`eve, Laboratoire d’Astrophysique
de l’Ecole Polytechnique F\'ed\'erale de Lausanne (EPFL),
1290 Sauverny, Switzerland}
\altaffiltext{11}{Dark Cosmology Centre, Niels Bohr Institute, University of Copenhagen, Juliane Maries Vej 30, DK--2100 Copenhagen {\O}, Denmark}
\altaffiltext{12}{Center for Astrophysics and Space Sciences
University of California, San Diego, 9500 Gilman Drive
La Jolla, California, 92093, jmoustakas@ucsd.edu}
\altaffiltext{13}{Department of Physics \& Astronomy, Western Washington University, Bellingham, WA 98225; kenneth.rines@wwu.edu}

\begin{abstract}
We present {\it Spitzer} MIPS 24\micron \ observations of 16 $0.4<z<0.8$ galaxy clusters
drawn from the ESO Distant Cluster Survey.
This is the first large 24\micron \ survey of clusters at intermediate redshift.  
The depth of our imaging corresponds to a total IR luminosity of $\rm 8\times10^{10}~L_\odot$, just below
the luminosity of luminous infrared galaxies (LIRGs), and $6^{+1}_{-1}$\% of 
$M_V < -19$ cluster members show 24\micron \ emission at or above this level.
We compare with a large sample of coeval field galaxies and find that 
while the fraction of cluster LIRGs lies significantly 
below that of the field, 
the IR luminosities of the field and cluster galaxies are consistent.  However, 
the stellar masses of the \edi \ LIRGs are systematically higher than those of the field LIRGs.  
A comparison with optical data reveals that $\sim$80\% of 
cluster LIRGs are blue and the remaining 20\% lie on the red sequence.
Of LIRGs with optical spectra, 88$^{+4}_{-5}$\% show [O~II] emission with EW([O~II])$>$5\AA,
and $\sim75$\% exhibit optical signatures of dusty starbursts. 
On average, the fraction of cluster LIRGs
increases with projected cluster-centric radius but remains systematically lower 
than the field fraction over the area probed ($< 1.5\times$\rtwo).  
The amount of obscured star formation declines 
significantly over the 2.4~Gyr interval spanned by the \edi \ sample, and the rate of decline
is the same for the cluster and field populations. 
Our results are consistent with an exponentially declining LIRG fraction, 
with the decline in the field delayed by $\sim1$ Gyr relative
to the clusters.

\end{abstract}


\section{Introduction}
\label{intro}
Rest-frame ultraviolet and optical studies indicate that the global
star formation rate (SFR) density has decreased by a factor of
approximately 10 between $z=0$ and $z=2$ 
({Gallego} {et~al.} 1995; {Lilly} {et~al.} 1996; {Madau} {et~al.} 1996; {Connolly} {et~al.} 1997; {Treyer} {et~al.} 1998; {Flores} {et~al.} 1999; {Steidel} {et~al.} 1999; {Wilson} {et~al.} 2002; {Giavalisco} {et~al.} 2004; {Hopkins} 2004).
What drives the decline in SFR?  Several scenarios have
been proposed, including the depletion of cold gas due to continuous
star formation and/or merger-driven bursts (e.g. {Bekki} \& {Couch} 2003); a decrease in the rate of
galaxy-galaxy interactions that trigger star formation (e.g. {Le F{\`e}vre} {et~al.}  2000; {Bridge} {et~al.}  2007; but see also {Lotz} {et~al.}  2008); and the
quenching of star formation in galaxies entering increasingly dense
environments as structure forms in the universe (e.g. {Larson} {et~al.} 1980).

Many of the mechanisms that fall within the above categories result
not only in a decline in the global SFR, but also in an environmental
dependence on SFR.  Indeed, it is well-established that the SFRs of
galaxies in the local universe correlate with environment, in the
sense that high-density environments have a lower fraction of
star-forming galaxies than low-density environments
(e.g. {Hashimoto} {et~al.} 1998; {Lewis} {et~al.} 2002; {G{\' o}mez} {et~al.} 2003; {Balogh} {et~al.} 2004; {Kauffmann} {et~al.} 2004; {Blanton} \& {Moustakas} 2009).
Large spectroscopic and photometric redshift surveys of the general
field have allowed the study of the SFR-density relation out to $z
\approx 1$.  Such studies have been undertaken in GOODS
({Elbaz} {et~al.} 2007), DEEP2 ({Gerke} {et~al.} 2007; {Cooper} {et~al.} 2007, 2008), VVDS
({Cucciati} {et~al.} 2006), COSMOS ({Cassata} {et~al.} 2007), and SHADES
({Serjeant} {et~al.} 2008).

A complementary approach to measuring the environmental dependence of star
formation has been the detailed study of galaxy clusters.  This method
ensures the inclusion of large numbers of galaxies in high-density
environments, which may be rare in field surveys.  Furthermore, cluster
studies can help clarify how the global cluster environment influences galaxy
properties relative to the local galaxy environment.  The SFRs in
clusters have been examined in various studies, including surveys of
multiple clusters such as CNOC
({Balogh} {et~al.} 1998, 1999; {Balogh} \& {Morris} 2000; {Ellingson} {et~al.} 2001), MORPHS
({Dressler} {et~al.} 1999; {Poggianti} {et~al.} 1999), \edi \ 
({Poggianti} {et~al.} 2006, 2008; {Finn} {et~al.} 2005), the \textit{Spitzer}/MIPS GTO team
({Bai} {et~al.} 2007; {Marcillac} {et~al.} 2007), and SMIRCS ({Saintonge} {et~al.} 2008; {Tran} {et~al.} 2009), 
as well as studies of individual clusters (e.g. {Finn} {et~al.} 2004; {Kodama} {et~al.} 2004; {Geach} {et~al.} 2006).

A full understanding of the evolution of the environmental dependence
of SFR has been impeded by at least two observational limitations.
First, most of the above general field and targeted cluster studies
were carried out in the rest-frame optical, which suffers from the
effects of dust extinction.  Because they make use of SFRs that are
computed from long-wavelength data that are less affected by dust, the
GOODS and SHADES studies are notable exceptions among the above
general field studies, as are the \textit{Spitzer}/MIPS GTO and SMIRC 
studies of clusters.  Second, the total number of clusters studied remains
fairly small in the face of the large cluster-to-cluster variations
observed (e.g. {Finn} {et~al.} 2005).  

This paper is one in a series based on the ESO Distant Cluster Survey
(EDisCS; {White} {et~al.} 2005; {Halliday} {et~al.} 2004; {Poggianti} {et~al.} 2006; {Desai} {et~al.} 2007).  
We address the above limitations by
presenting \textit{Spitzer}/MIPS observations of 16
intermediate-redshift ($0.42 < z < 0.8$) EDisCS clusters.  Not only
does this triple 
the number of well-studied clusters
at these redshifts, but the MIPS observations allow us to characterize
the dust-obscured SFRs.  These observations also have the advantage
that they provide us with a SFR-limited sample of cluster galaxies.  

This paper is organized as follows.  
We describe the cluster sample in \S\ref{ediscs} and the 
\textit{Spitzer}/MIPS
data in \S\ref{data}.  We then describe the selection of the cluster members in 
\S\ref{samples} and the conversion from observed 24\micron \ flux to total
IR luminosity in \S\ref{lirtot}.  In \S\ref{gems}, we describe the sample
we use for a field comparison. We present our results in \S\ref{results}, 
including the spatial and
luminosity distributions of MIPS galaxies.  In
\S\ref{discussion}, we discuss our results in the context of cluster evolution scenarios, 
and we present our conclusions in \S\ref{conclusions}.

We adopt a $\Lambda$CDM cosmology, 
assuming $\Omega_0 = 0.3$, $\Omega_\Lambda = 0.7$, and $\rm {H_0} = {70~km~s^{-1}~Mpc^{-1}}$ unless
otherwise noted.  All magnitudes are relative to Vega.

\section{ESO Distant Cluster Survey}
\label{ediscs}
The clusters targeted in this survey are drawn from EDisCS
({White} {et~al.} 2005).  
EDisCS is an ESO Large Programme that targeted 20 fields
with VLT imaging and spectroscopy, and NTT near-IR
imaging.  In those fields, 26 structures (groups and clusters) 
have been identified ({Halliday} {et~al.} 2004; {Milvang-Jensen} {et~al.} 2008).
One field contained no signficant structure, and the structure cl1103.7-1245 at 
$z=0.96$ does
not contain many spectroscopically-confirmed members.
The remaining 18 fields contain 17 primary structures and 7 secondary ones, 
where primary denotes the most massive structure in each field. 
These comprise the only sizable sample of low-mass clusters
that has been studied in such detail at high redshift.  
Mass estimates for the \edi \ clusters have been derived from weak lensing ({Clowe} {et~al.} 2006)
and velocity dispersions.

We targeted only 16 primary clusters with $Spitzer$ (we omit the 
primary structure in the cl1119.3-1129 field because of its low velocity dispersion).
The redshift and velocity dispersions of the clusters are listed in Table \ref{detect}.
All 16 clusters have extensive ground-based data
that cover the same approximate area imaged by {\it Spitzer}, including
multiband photometry and spectroscopy.  
There are 30-50 spectroscopically-confirmed members per
cluster.  
Photometric redshifts, spectroscopic
results, and ground-based morphologies are available for a 
5\arcmin$\times$5\arcmin \ region around each cluster.  
Furthermore, 
derived-data products such as k-corrected absolute magnitudes and estimates 
of the total number of member galaxies ({Pello} {et~al.} 2009; {Rudnick} {et~al.} 2009) are readily
available.  
The velocity dispersions and redshifts of the \edi \ clusters are not correlated
(see Table \ref{detect}), 
an important point to demonstrate before looking
for evolutionary trends within the sample.

\section{Spitzer/MIPS Data}
\label{data}
Infrared observations of cluster galaxies by the {\it Infrared
Space Observatory} (ISO; {Kessler} {et~al.} 1996) 
allowed the first look at
obscured star-forming galaxies in distant clusters ({Metcalfe} {et~al.} 2005).  
However,
systematic coverage of clusters could not be conducted with {\it ISO} given the
time it would have required to obtain multi-positioned and heavily-overlapped
rasters of such targets.  The {\it Spitzer Space Telescope} ({Werner} {et~al.} 2004) 
provides not only the desired coverage of clusters, but with the
improved sensitivity necessary to probe the infrared properties of cluster
galaxies to lower masses and to larger redshifts, providing a unique
opportunity to explore their evolution.

\subsection{Observations}
We obtained images of the clusters at 24\,$\micron$ using the Multiband
Imaging Photometer for {\it Spitzer} (MIPS; {Rieke} {et~al.} 2004).  Observations of the
16 clusters were taken during Cycles 2 and 3 under guest observer programs 20009 and 30102.
We imaged the central 5\arcmin$\times$5\arcmin \ of each
cluster to match the areal coverage of the ground-based VLT data. 
This area corresponds to a projected size
of 1.8$\times$1.8 Mpc at z=0.5 and 2.3$\times$2.3 Mpc at z=0.8, which incorporates
$>90$\% of the volume within the virial region at both epochs, assuming a 
typical comoving virial radius
of 1~Mpc.

To map the 5\arcmin$\times$5\arcmin \ area, 
we use MIPS Photometry
in large field mode.   We 
complete 10 cycles of 10 sec exposures for the $0.42 < z < 0.52$ clusters and 20 cycles
of 10 sec exposures for the higher-redshift clusters.
We use a 20\arcsec \ sky
offset, which effectively doubles our on-source exposure time.

\subsection{Data Reduction}

The post-basic calibration data images produced by the {\it Spitzer} Science Center (SSC) 
pipeline show large scale variations in the sky level
that are not effectively corrected by the flat field.  We therefore start our data
reduction by applying an additional flatfield correction to the basic calibration data (BCD)
images.  To do this, we first run SExtractor ({Bertin} \& {Arnouts} 1996) on individual BCD images.  
Using the source list from SExtractor, we mask all pixels that lie
within a 10-pixel radius of an object.  We then average the masked BCD images using 
the IRAF routine imcombine to create a flat. 

We use the MOsaicker and Point source EXtractor (MOPEX\footnote{APEX was written for the SSC by David Makovoz.}) 
routine {\it mosaic.pl} to 
create a mosaic of the flattened images, and we 
fit the point response function (PRF) in each mosaiced image using the Astronomical Point Source
EXtraction (APEX) routine {\it prf\_estimate.pl}.
We then use the APEX routine {\it apex\_1frame.pl} to identify sources and 
extract aperture photometry.  We use the total flux value calculated by MOPEX.

To test for systematics in our data reduction procedure, we compare the aperture fluxes from the 
mosaic made from MOPEX with 
a mosaic made using the MIPS Instrument Team's Data Analysis
Tool v. 3.06 (DAT; {Gordon} {et~al.} 2005) for the cluster CL1216.  The average ratio of the 
DAT to MOPEX aperture 
fluxes measured in an aperture with a 4-pixel (10.2\arcsec) diameter is $1.02 \pm 0.09$, 
so fluxes are consistent at the 10\% level.  The RMS increases with aperture
size; the average ratio is  
$1.05 \pm 0.18$ using an aperture with a diameter of 6 pixels (15.3\arcsec).
Fluxes resulting from MOPEX and DAT reductions are consistent, and we use the MOPEX reduction
for the rest of the sample.  

\subsection{Completeness}
We use the IRAF\footnote{IRAF is distributed by the National Optical Astronomy Observatory, which is operated by the Association of Universities for Research in Astronomy (AURA) under cooperative agreement with the National Science Foundation.} artdata package to estimate our detection efficiency as a function of 
source brightness.  The cluster galaxies that we are studying are smaller than the
MIPS 24\micron \ PSF, so we add point sources into the final mosaiced image.
We add 1000 sources with fluxes ranging uniformly from 10$\mu$Jy to 180$\mu$Jy.
The artificial sources are positioned randomly on each mosaiced image, 
avoiding edges and previously-placed artificial sources.  We limit the number of
artificial sources to 10 per image so that we do not significantly alter 
the source density and repeat the simulation 100 times per cluster image to accumulate
1000 artificial sources per cluster.
We convolve the sources with a PRF from the SSC derived from a mosaiced 24\micron \ image.
We use this rather than the PRF measured from each image because our measured PRFs are
frequently contaminated by nearby sources whereas the SSC PRF is not.
We then rerun {\it apex\_1frame.pl} and determine the fraction of 
artificial sources detected as a function
of source brightness.  The results show that our 80\% completeness limit
ranges from 72$\mu$Jy to 100$\mu$Jy, with the 80\% completeness flux level listed for 
individual clusters in Table \ref{detect}.

\section{Assembly of Cluster Samples}
\label{samples}
\subsection{Optical Counterparts of 24\micron \ Sources \label{optcount}}

Although the alignment of the optical and 24\micron \ images is good, we find systematic offsets between the images that can be
as large as 1.1\arcsec.  To correct for this misalignment, we perform a first-pass match between the optical and IR sources.  
From the matched sources, we calculate the
average offset between the optical and 24\micron \ positions and then adjust the 24\micron \ coordinates so that 
the average offset between the optical and IR positions is zero in both RA and Dec.  We then rematch the optical 
and IR sources using the shifted 24\micron \ coordinates.

We find a total of 2337 24\micron \ sources in the 16 cluster images that have a signal-to-noise ratio greater than 2.5.
We are able to match
1911 of these ($82\pm1$\%) to optical counterparts in the \edi \ source catalogs using a match radius
of 2\arcsec \ between the 24\micron \ and optical source.
When multiple sources lie within 2\arcsec, we select the optical counterpart that is closest to
the 24\micron \ source.  Overall, 222 sources (9\%) have more than one optical match within 2\arcsec.

We find 426 24\micron \ detections that are not matched to an object in the \edi \ catalog.
Some of these IR sources (79) are false detections, lying near the edge of the
24\micron \ image or associated with the Airy ring of a bright 24\micron \ source.
Other IR sources (48) are unmatched because they 
overlap a bright, extended optical source, and thus any optical counterpart is
undetectable in the optical image.
The number of remaining unmatched 24\micron \ sources 
is 268.  Of these, 125 sources (5\% of the IR galaxies) 
appear to be a blend of one or more
optical sources; these have an optical counterpart, it just is not clear which one is the counterpart.  
We consider the remaining 143 sources to be obscured sources; 
76 coincide with a faint optical source that is below the detection
limit of the \edi \ catalog, and the remaining 67 sources have no optical counterpart in the 
\edi \ I-band image.
These sources are likely to lie at redshifts beyond
our clusters (e.g. {Le Floc'h} {et~al.} 2005).

\subsection{Selection of Cluster Members}

We calculate cluster membership in two complementary ways.  
We use spectroscopic redshifts 
to conclusively establish the membership when available.  The majority of
galaxies in each cluster, however, does not have spectroscopy, and we
therefore use photometric redshifts 
to determine cluster membership for the galaxies with no
spectroscopy.  The EDisCS spectroscopy is described in detail
in {Halliday} {et~al.} (2004) and {Milvang-Jensen} {et~al.} (2008), and the photometric
membership techniques are described in detail in {Pello} {et~al.} (2009)
and {Rudnick} {et~al.} (2009).  Here we provide a brief summary of the
techniques used in this paper.

Spectroscopic members are those with velocities within $\pm3\sigma$
from the cluster redshift.  Spectroscopic membership information
supercedes all photometric information.  The photometric redshifts
have been computed from the full optical/NIR photometry and have been
calibrated from the extensive EDisCS spectroscopy to cull non-members,
while retaining $>90\%$ of all confirmed cluster members independent
of rest-frame color, down to the spectroscopic magnitude selection
limit, which was $22<I<23$ depending on the cluster. The photometric
redshifts only yield robust membership classifications when optical
and NIR data are present, and so we limit our sample to those areas of
each field with adequate exposure in all bands.  

To test the
reliability of our photometric redshifts for the IR-selected galaxies
in our sample, we compare the completeness and the contamination of the
photometric membership for IR and non-IR detected galaxies using our
extensive spectroscopy.  We define the completeness as the number of
spectroscopic members that are also photometric redshift members,
divided by the number of spectroscopic members. 
We find a completeness of $85 \pm 6\%$. For the subsample of
spectroscopic members that are also 24$\mu m$ sources, we find a
completeness of $85\pm12\%$.  We define contamination as the number of
spectroscopic non-members which are classified as photometric members,
divided by the total number of spectroscopic members that are classified as 
photometric members. This yields a contamination of
$48 \pm 3\%$. For the subsample of spectroscopic members that are also
24$\mu m$ sources, we find a contamination of $53 \pm 6\%$. Thus, the
completeness and contamination of the 24$\mu m$ sources are entirely
consistent with the optically selected cluster sample down to the
spectroscopic magnitude limit.  
{Marcillac} {et~al.} (2007) also find that the accuracy of photometric redshifts
for IR-selected galaxies in a $z=0.83$ cluster is the same as that
for all cluster members.

\section{Total Infrared Luminosity}
\label{lirtot}
We use the {Dale} \& {Helou} (2002) models to estimate total IR luminosity (3-1100 \micron) 
from 24\micron \ fluxes
as a function of galaxy redshift.
To first calculate the 24\micron \ luminosity, we multiply the 24\micron \ flux (in Jy) 
by $4 \pi d_L^2$, where
$d_L$ is the luminosity distance corresponding to the cluster redshift, and by 
$c/23.8$\micron,
the central frequency of the 24\micron \ bandpass.
  We scale the 24\micron \ luminosity by the conversion found from the {Dale} \& {Helou} (2002) 
templates 
corresponding to the cluster redshift to estimate \lir.
According to the {Dale} \& {Helou} (2002) templates, 
the error associated with estimating the IR luminosity solely from the observed 
24um flux varies with redshift; for the redshift range spanned by the \edi \ clusters, 
the error ranges from a minimum of 5\% at $z=0.6$ to a maximum of 22\% at $z=0.8$.  
Finally, 
we divide \lir \ by the luminosity of the Sun, where 
$\rm L_\odot = 3.826 \times 10^{33}~ergs~s^{-1}$.

Active galactic nuclei (AGN) will contaminate our measurements with flux that
is not associated with star formation.  Mid-infrared colors from the Infrared
Array Camera (IRAC; {Fazio} {et~al.} 2004), spanning 3-8\,$\micron$, will help
identify AGN ({Stern} {et~al.} 2005; {Donley} {et~al.} 2007; {Lacey} {et~al.} 2008) and 
will be presented in a future paper.  
However, {S{\'a}nchez-Bl{\'a}zquez} {et~al.} (2009) analyze the optical spectra of \edi \ galaxies and 
find that most of the emission-line galaxies are powered by star-formation rather
than AGN.  
Furthermore, {Bell} {et~al.} (2005) estimate
that $\lesssim 15$\% of the total infrared luminosity density at 
$0.65<z<0.75$ is
from sources with significant AGN emission, and the contribution is likely much
lower since the infrared luminosity in these galaxies may also arise from star
formation (see also {Robaina} {et~al.}  2009 for a more detailed 
discussion).  Similarly, the {Bai} {et~al.} (2007) and {Marcillac} {et~al.} (2007)
studies of clusters at similar redshifts to our sample find that only $\sim$4 out
of 66 IR-detected galaxies are unambiguously AGN and argue that the
rest of their galaxies are dominated by dusty starbursts.  
Finally, {Geach} {et~al.} (2009) observe a sample of 12 galaxies that are members
of a $z = 0.4$ cluster with the Infrared Spectrograph (IRS; {Houck} {et~al.} 2004).  
The results of the IRS analysis indicate 
that the mid-IR 
emission for 11 of 12 galaxies is powered by a starburst rather than AGN.
Based on these
results, we conclude that contamination by AGN is likely to be small.

We use the relation from {Kennicutt} (1998) to convert from \lir \  (8-1000 \micron) 
to SFR.  There is a slight discrepancy between the definition of \lir \ used by the
{Dale} \& {Helou} (2002) models (3-1100\micron) and that used in the Kennicutt star-formation 
conversion (8-1000\micron), but this impacts the inferred SFR by less than 5\% (D. Dale 2009, private communication).  

When we translate the 80\% completeness flux limits listed in Table \ref{detect} to 
IR luminosities, 
we find that we do not probe as deeply in the
higher-redshift clusters.
CL1216 and CL1353 have the highest 80\% completeness luminosities of 
$\rm log_{10}(L_{IR}/L_\odot) = 10.91$, 
which corresponds to an IR-derived SFR of $\sim13$~\smy.
This is just slightly below
the luminosity of luminous infrared galaxies (LIRGs; $\rm log_{10}(L_{IR}/L_\odot) > 11, \  L_{IR} > 3.8 \times 10^{44}~ergs~s^{-1}$) ({Sanders} \& {Mirabel} 1996). 
Therefore, we are sampling the population
of luminous infrared galaxies rather than normal star-forming 
galaxies uniformly across the redshift range.
Because our \lir \ limit is so close to the LIRG limit and given the uncertainties in
computing \lir \ solely from observed 24\micron \ flux, we refer to our IR galaxies as
LIRGs.

\section{GEMS Field Sample}
\label{gems}

Discrimination among cluster-specific processes that might affect
the gas content of member galaxies requires a comparison
between the star-formation
properties of cluster and field  galaxies.  
The Galaxy Evolution from Morphology and SEDs project (GEMS; {Rix} {et~al.} 2004) 
provides a good field sample to compare with our clusters because it is
currently the largest sample of intermediate-redshift galaxies with
accurate photometric redshifts and space-based morphologies.
The GEMS survey covers an $\sim$800 square arcminute region centered
on the Extended Chandra Deep Field S (ECDF-S).  
The GEMS survey area was chosen to
overlap with the Classifying Objects by Medium-Band Observations
in 17 filters survey (COMBO-17; {Wolf} {et~al.} 2001, 2004), which provides
accurate photometric redshifts ($\sigma_z/(1+z)\sim 0.02$) out to 
$z < 1.2$.  
Most importantly for our purposes, the GEMS survey area was also
imaged by the $Spitzer$ MIPS GTO team at 24\micron \ ({Papovich} {et~al.} 2004)
to a depth comparable to our imaging ($5\sigma$ detection limit $\rm = 83~\mu Jy$).  
The IR properties of the
GEMS galaxies are studied in detail by {Bell} {et~al.} (2005) and {Le Floc'h} {et~al.} (2005).  

To build a comparison sample for our clusters, we select all GEMS
galaxies within $0.42 < z < 0.8$, $I < 24$, and that lie within the 24\micron \ imaging area.  
In addition, we require a signal-to-noise ratio of at least 3 in both $I$ and $V$ observed
magnitudes.
We compute \lir \ 
from the observed 24\micron \ flux 
using the redshift-dependent conversion described in \S\ref{lirtot} rather than
using the \lir \ values of {Bell} {et~al.} (2005).
When comparing with \edi \ galaxies, we impose the following \lir  \ cuts:
$\rm log_{10}(L_{IR}/L_\odot) > 10.75$ for the $0.42 < z < 0.6$ galaxies, and 
$\rm log_{10}(L_{IR}/L_\odot) > 10.95$ for the $0.6 < z < 0.8$ galaxies.  The \lir \ 
limit for the low-$z$ sample is set by the depth of the low-$z$ \edi \ imaging, whereas
the \lir \ limit for the high-$z$ sample is set by the depth of the GEMS imaging.

\section{Analysis \& Results}
 \label{results}

\subsection{Properties of IR Galaxies}
\subsubsection{IR Luminosity Distribution}
 \label{lirdist}
\begin{figure*}[h]
\plottwo{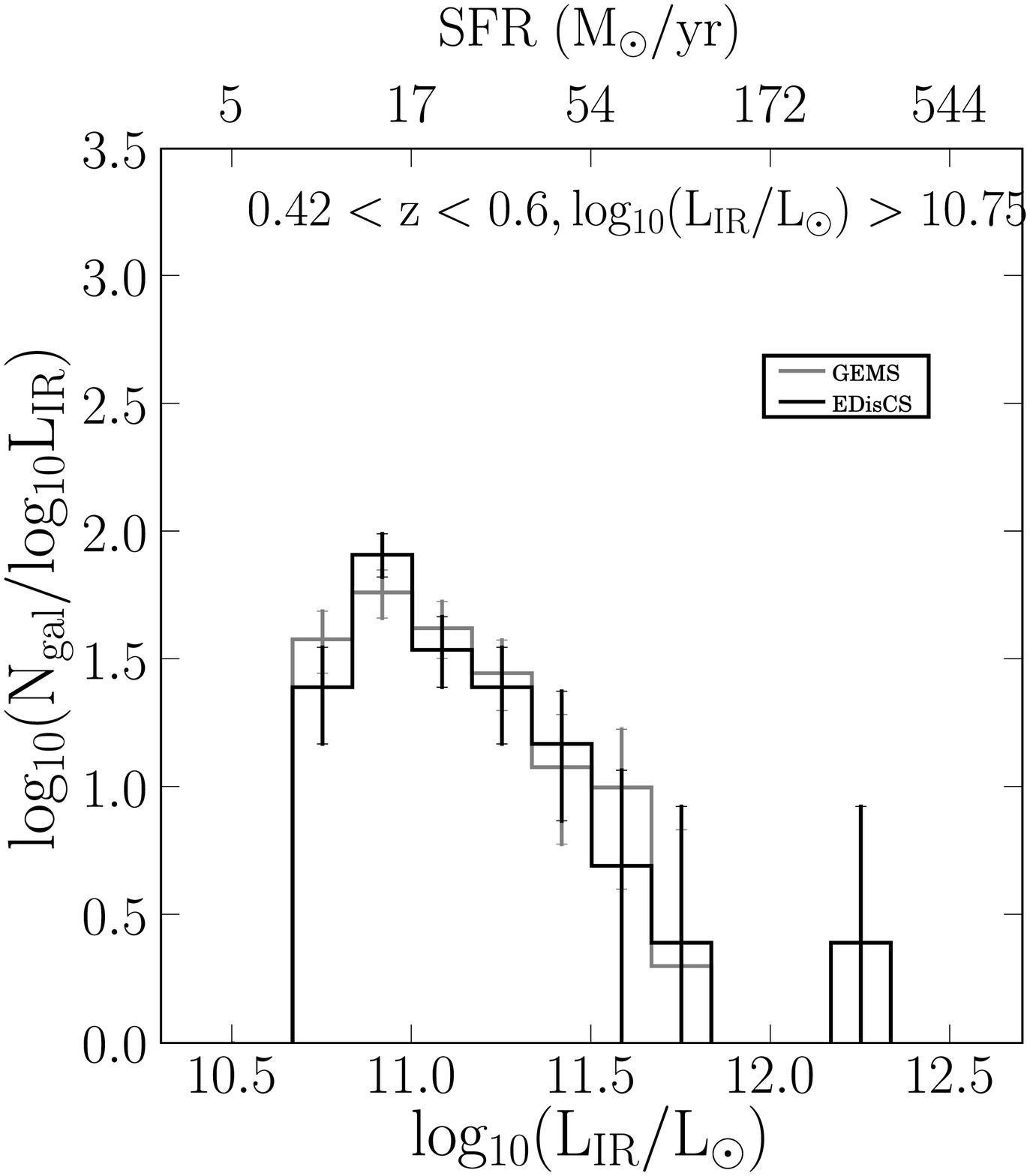}{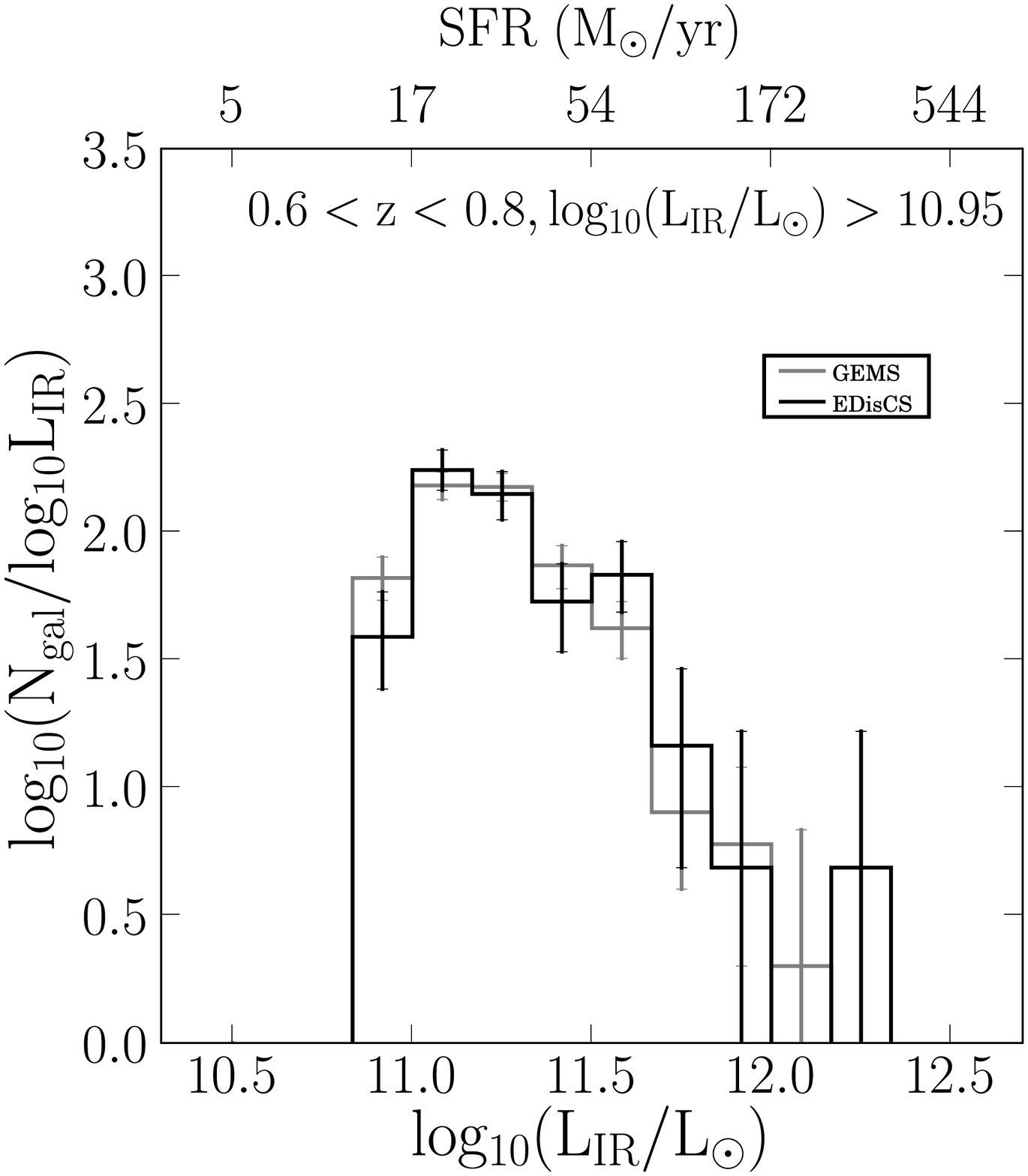}
\caption{Distribution of \lir \ for \edi \ (black) and GEMS (gray) samples for the 
$0.42 < z < 0.6$ sample (left) and $0.6 < z < 0.8$ sample (right).  A lower IR luminosity cut is used
for the lower-redshift sample, but in each redshift bin, 
the same \lir \ and magnitude cut is applied to the \edi \ and 
GEMS galaxies.  The 
\edi \ histograms are scaled to match the total number of GEMS galaxies.
The errorbars show Poisson errors. The IR luminosities of the field and clusters
galaxies are indistinguishable for both low and high-$z$ samples, indicating that the \lir \
distribution of the most actively star-forming galaxies is not affected by the cluster environment.  
The top horizontal axes show IR-derived SFRs.
\label{gemscomp}}
\end{figure*}

We compare the \lir \ distribution of our full sample with that of the GEMS 
field sample in Figure \ref{gemscomp}.  
We split both the field and cluster samples at $z = 0.6$ to minimize
evolutionary effects.
The $z < 0.6$ sample includes 87 GEMS galaxies
and 75 \edi \ galaxies, and
the $z > 0.6$ sample includes 250 GEMS galaxies and 102 \edi \ galaxies.
We show the luminosity distribution in terms of $\rm N_{gal}/log_{10}L_{IR}$
in Figure \ref{gemscomp} for the lower (left) and
higher-redshift (right) samples.  
After scaling the \edi \ distribution and errors to adjust for the difference 
in sample size, the GEMS and \edi \ distributions 
agree within errors for both the low and high-redshift samples.  
Furthermore, a two-sided 
Kolmogorov-Smirnoff (KS) test
can not distinguish the cluster and field distributions in both redshift bins.  
Thus, the IR luminosity distribution 
of the most actively star-forming galaxies is not affected by the cluster environment.

\subsubsection{Colors of IR Galaxies \label{optir}}
\begin{figure*}[h]
\epsscale{0.9}
\plotone{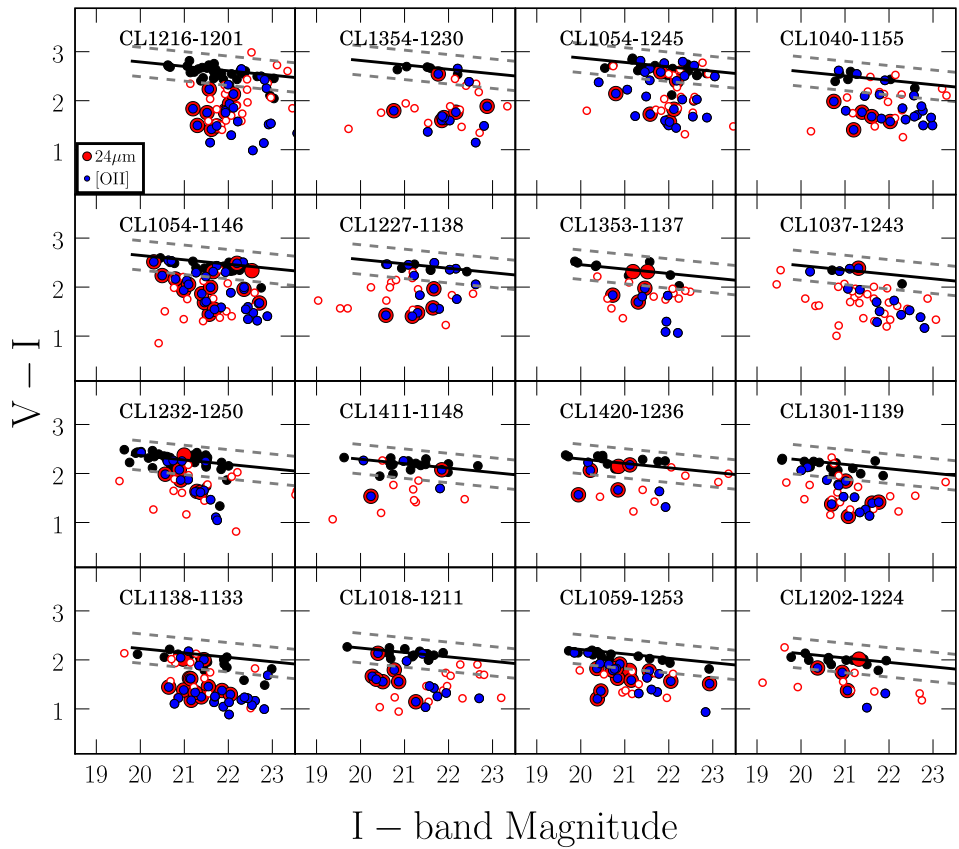}
\caption{$V-I$ color versus $I$-band magnitude for all clusters.  
The red symbols denote galaxies
with 24\micron \ emission, the blue symbols show galaxies with [O~II] emission in
their spectra ({Halliday} {et~al.} 2004; {Milvang-Jensen} {et~al.} 2008), and the black symbols represent the remaining 
spectroscopic members.  The open red symbols show photoz members with 24\micron \
emission.
The solid lines in each panel show the fit to the 
red sequence, and the dashed lines mark a color offset of $\Delta(V-I)< 0.3$ from the red
sequence, illustrating the selection criteria for red sequence members as presented in {De Lucia} {et~al.} (2006).  
The majority (80\%) of the MIPS sources lie blueward of the red sequence.  \label{cmda}}
\end{figure*}

The location of the 24\micron \ sources on the cluster 
color-magnitude diagrams is shown in Figure
\ref{cmda}.  The solid black line in each panel is the fit to the red sequence from
{De Lucia} {et~al.} (2007), who 
assume a fixed red-sequence slope of $-0.09$ and fit the zeropoint to the non-emission line
spectroscopically-confirmed cluster members.
The dashed lines mark a color offset of $\Delta(V-I)< 0.3$ from the red
sequence, illustrating the selection criteria for red sequence members used here 
and in previous analyses of the \edi \ clusters ({De Lucia} {et~al.} 2007; {S{\'a}nchez-Bl{\'a}zquez} {et~al.} 2009; {Rudnick} {et~al.} 2009).
We define blue cluster galaxies as those with $V-I$ colors at least 0.3
magnitudes bluer than the red sequence.
One main result from Figure \ref{cmda} is that $21^{+6}_{-4}$\% (16/77)
of the spectroscopically-confirmed cluster LIRGs lie on the
red sequence while the remaining $79^{+4}_{-6}$\% (61/77) lie in the blue cloud.  This fraction of
red LIRGs is higher than that observed by {Tran} {et~al.} (2009) in a $z = 0.35$ cluster and is comparable to 
the fraction they measure in the field.  {Tran} {et~al.} (2009) use a $B-V$ cut to select red galaxies rather than
the $V-I$ cut used in this paper, and 
this might account for some of the discrepancy.

Figure \ref{cmda} shows that there is not a one-to-one correspondence between [O~II]-emitting 
(blue circles) and IR-bright galaxies (red circles).  
Only a minority ($29\pm3$\%) of spectroscopic members 
with [O~II] emission (EW(O~II)$>5$\AA) are detected at 24\micron.
Conversely, 
83$^{+4}_{-5}$\% (64/77) of the IR galaxies show [O~II] emission
in their optical spectra (EW([O~II])$>$5\AA).  The 
remaining IR galaxies have weaker emission
lines and lie predominantly on the red sequence, consistent with the 
results of {S{\'a}nchez-Bl{\'a}zquez} {et~al.} (2009).  
The 13 IR sources with no [O~II] emission do not dominate the
total \lir \ but contribute a fraction that is entirely consistent 
with their number.
Thus, we do not find a large fraction of optically-selected cluster
members whose star-formation is completely obscured at visible wavelengths.  
However, the IR-derived 
SFRs greatly exceed those derived from 
dust-corrected [O~II] emission for the majority of the IR galaxies;
the median ratio of SFR(IR)/SFR(O~II)
is 2.9 for the $Spitzer$-detected galaxies in the \edi \ clusters
({Vulcani} {et~al.} 2010).  

Of particular interest are the 24\micron \ sources on the red sequence.  
One possibility
is that they are ellipticals with AGN rather than star-forming galaxies.  
Of the red sequence galaxies for which we have $HST$ morphologies, 
6 have IR emission: one is an elliptical, and the 
remaining 5 are normal spirals.  Thus, the majority of the reddest 
IR sources appear to be red because of dust, and future analysis will 
probe the properties of members on or near the red sequence, 
including AGN emission, in more detail.

\subsubsection{Magnitudes of IR Galaxies}
\begin{figure*}
\plotone{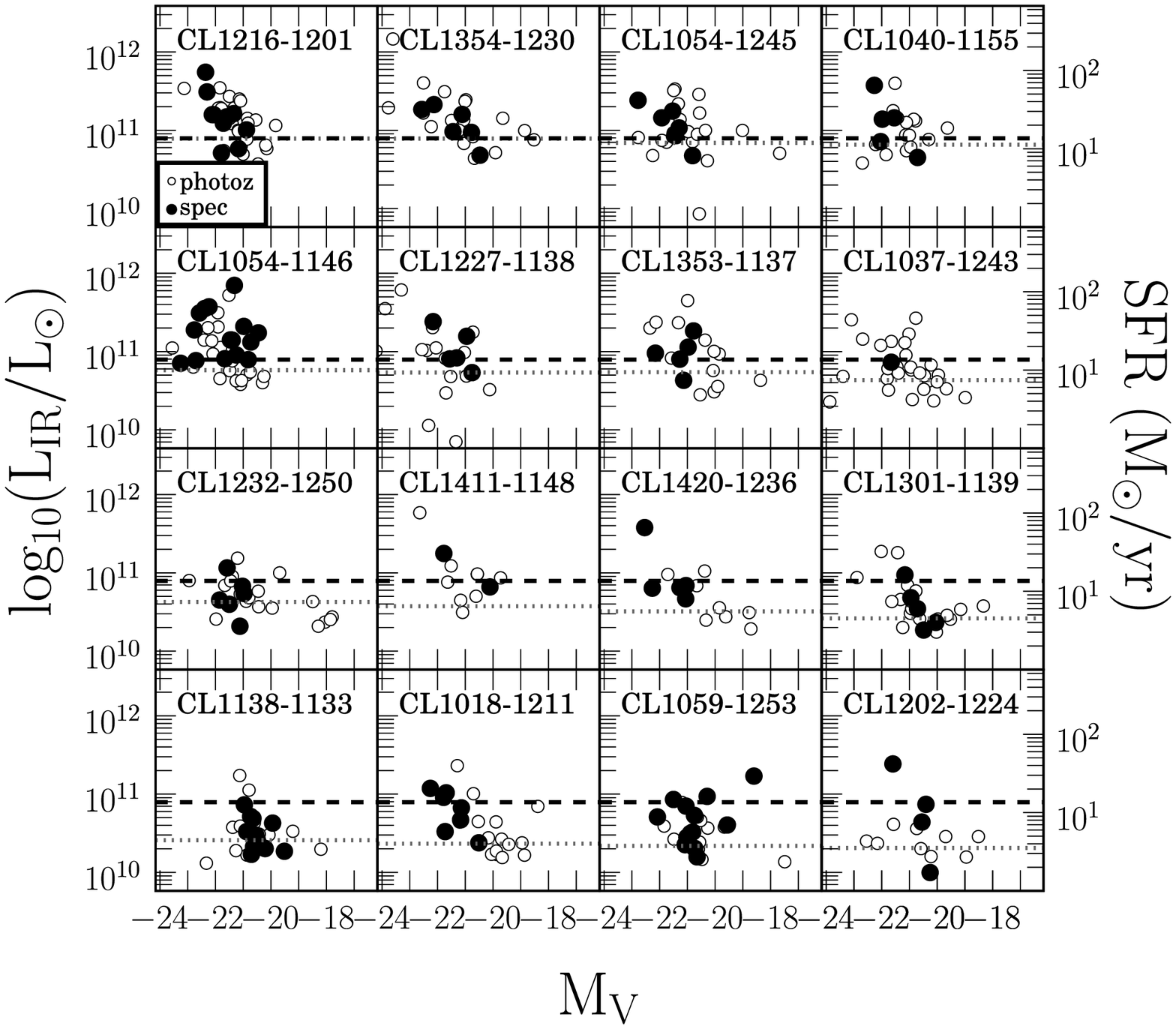}
\caption{\lir \ versus $\rm M_V$ for \edi \ clusters.  The filled and open circles show the spectroscopic
and photometric-redshift members, respectively.  The dotted horizontal line shows the 
80\% completeness limit for each cluster.  
The dashed horizontal line shows $\rm log_{10}(L_{IR}) = 10.91$, the 80\% completeness limit of
CL1216.  The cluster name is listed at the
top of each panel, and the clusters are ordered by
decreasing redshift from left to right and top to bottom.  
The  \lir \ of the 24\micron-emitting galaxies decreases with redshift.  
The right-hand vertical axes show IR-derived SFRs.
\label{lirMra}}
\end{figure*}

In Figure \ref{lirMra} we show \lir \ versus $\rm M_V$ for
individual clusters, with the panels ordered by decreasing redshift
from left to right and top to bottom.  
The dotted line shows the 80\% completeness limit for each cluster.  
The dashed line shows the 80\% completeness limit for CL1216, 
which has the highest threshold of all the clusters.  
The cluster-to-cluster variations in the distribution of \lir \ are striking and illustrate
the need for large samples of clusters to properly characterize 
star formation in dense environments at a given
epoch. Furthermore, the higher-redshift clusters appear to have more galaxies with high
values of \lir.  
The redshift range of the \edi \ clusters corresponds to a time interval of 2.4~Gyr.
Given the dramatic decline in SFRs since $z\sim 1$ as discussed in \S\ref{intro}, 
one might expect to observe evolution within the \edi \ sample, and we examine
this is more detail in \S\ref{zevol}.

\begin{figure*}
\plotone{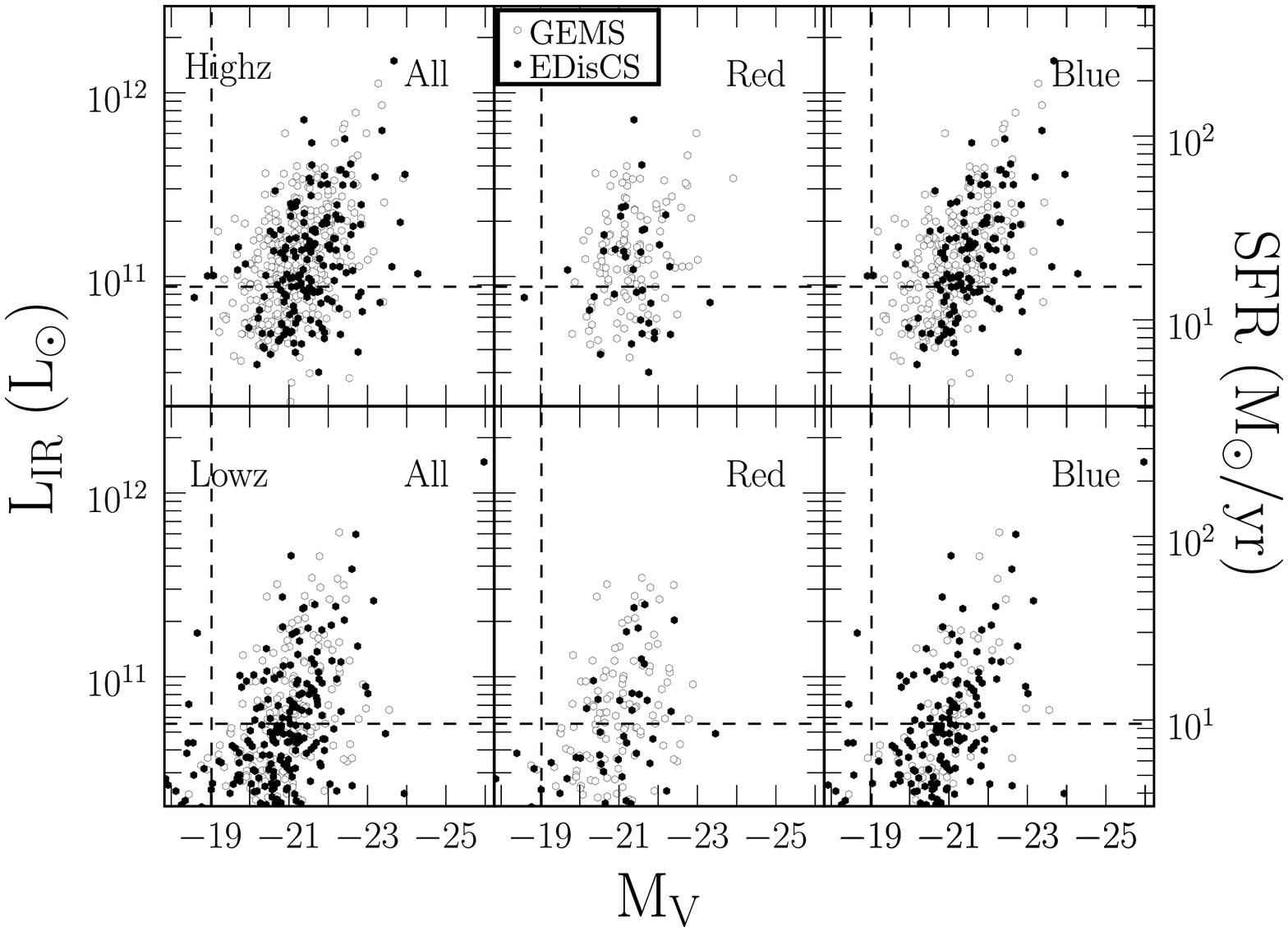}
\caption{IR luminosity versus rest-frame absolute V magnitude.  The top and bottom rows show the 
$0.6 < z < 0.8$ and $0.42 < z < 0.6$ \edi \ (black) and GEMS (white) galaxies, respectively.  
The columns show the \lir-$M_V$ distribution for all, red, 
and blue galaxies, respectively.  The dashed horizontal lines show the IR completeness limit 
while the dashed vertical lines show the magnitude limit of $M_V < -19$.  The high-z blue cluster 
galaxies have systematically brighter luminosities than the high-z blue field galaxies.  
The \lir-$M_V$ distributon of all 
other subgroups are consistent. \label{lirmvredblue}}
\end{figure*}

In Figure \ref{lirmvredblue}, we combine the $0.42 < z < 0.6$ and $0.6 < z < 0.8$ cluster 
samples to show IR luminosity versus rest-frame absolute V magitude as a function of galaxy color.  
The corresponding GEMS samples are shown with gray circles.  
Both field and cluster samples are complete in the top right quandrant of each plot, 
at $M_V > -19$ (set by the depth of the GEMS survey) and $log_{10}(L_{IR}) > log_{10}(L_{lim})$, 
where $log_{10}(L_{lim})=10.95$ for the high-z samples and 10.75 for the lower-$z$ samples. 
For the \edi \ clusters, we separate the blue and red galaxies as described in \S\ref{optir}.
To separate red and blue GEMS galaxies in
an analogous manner, we fit the red-sequence zeropoint of the \edi \ clusters 
as a function of cluster redshift.  The resulting red sequence is given by:
$-0.09 (I-20) + 1.9(z-0.42) +2.14$.  
As with the \edi \ galaxies, we define blue galaxies as those with $V-I$ colors at least 0.3
magnitudes bluer than the red sequence.  
The main result from Figure \ref{lirmvredblue} is that 
the $L_{IR}-M_V$ distributon in all subplots are consistent except for the high-z blue galaxies;  
the high-z blue cluster galaxies have systematically brighter luminosities than the high-z blue field galaxies.

\subsubsection{Stellar Masses of IR Galaxies}
\begin{figure*}
\plotone{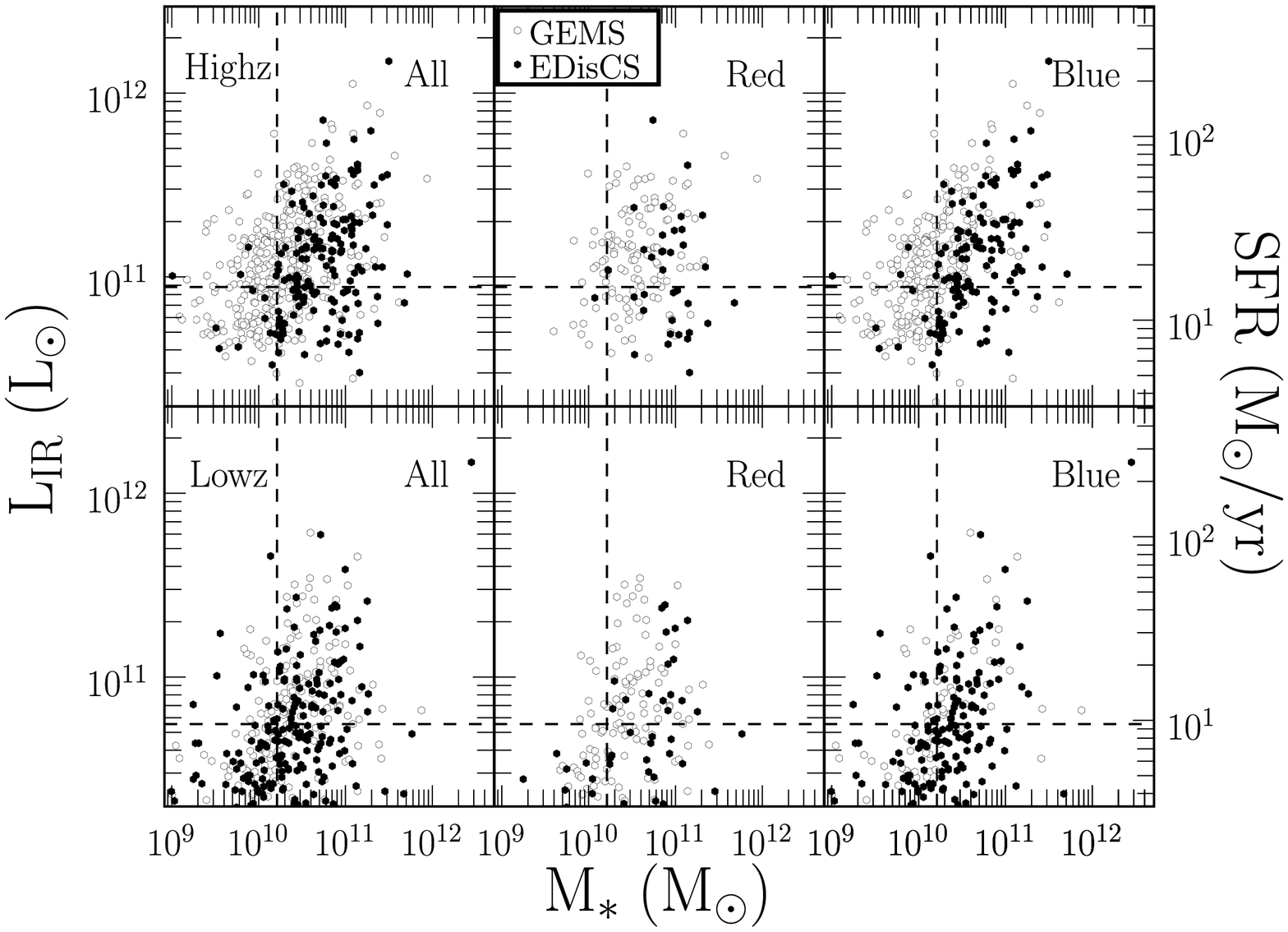}
\caption{IR luminosity versus stellar mass.  The top and bottom rows show the $0.6 < z < 0.8$ and $0.42 < z < 0.6$ \edi \ (black) and 
GEMS (white) galaxies, respectively.  The dashed horizontal lines show the IR completeness limit while the dashed 
vertical line shows the magnitude limit of $M_V < -19$.  The EDisCS IR galaxies have higher stellar masses on 
average than the GEMS galaxies. \label{lirstellmassredblue}}
\end{figure*}

In Figure \ref{lirstellmassredblue}, we show IR luminosity versus stellar mass for 
the higher-$z$ and lower-$z$ samples.  Again, we show the blue and red galaxies separately.  
We calculate stellar mass using the relation from {Bell} {et~al.} (2003):  
\begin{equation}
log_{10}(M_*/L)_B = 1.737(B-V)-0.942.
\end{equation}  
We adopt a stellar mass completeness limit of $log_{10}(M_*) = 10.2$ (vertical dashed line), 
which we calculate using the magnitude limit $M_V = -19$ and an assumed color of $B-V < 1$.  

A two-sided, two-sample Kolmogorov-Smirnov test indicates that the $L_{IR}-M_*$ distribution 
of the higher-$z$ galaxies are significantly different (where we define
significant as $>3\sigma$).
As shown in Figure \ref{gemscomp}, the IR luminosities of the field and cluster galaxies are consistent.  
The $L_{IR}-M_*$ distributions differ because 
the stellar masses of the \edi \ red and blue IR galaxies are systematically higher than the stellar masses of the GEMS galaxies.  
In the lower-$z$ samples, only the $L_{IR}-M_*$ distribution of the red galaxies differs
significantly, and the sense of the difference is the same:  the red \edi \ galaxies have higher stellar masses 
on average while their IR luminosities 
are consistent with the GEMS galaxies.

The results imply that the specific star-formation rates of all higher-$z$ 
cluster galaxies and the red lower-$z$ cluster galaxies are lower than the corresponding field
galaxies.  
Given the uncertainties associated with comparing the colors and magnitudes from two different 
surveys, the significance of the difference
in the $L_{IR}-M_*$ distributions is difficult to assess.  
However, {Vulcani} {et~al.} (2010) perform a similar analysis, comparing the spectroscopically-confirmed \edi \ galaxies to field galaxies from the 
All-Wavelength Extended Groth Strip International Survey (AEGIS) ({Noeske} {et~al.} 2007).  They measure star-formation rates from both [O~II] emission and
24\micron \ emission, and they find that both the lower and higher-$z$ cluster 
galaxies have lower SFRs at a given stellar mass compared to the field.  
When comparing only blue galaxies, they find that the $z > 0.6$ cluster galaxies have lower specific SFRs while
the $z < 0.6$ blue cluster galaxies do not.  The difference becomes statistically significant in both redshift bins
when red galaxies are included.  In this paper we find similar results using a different field sample 
and a different method of measuring star-formation rates.  The addition of [O~II] emission allows 
{Vulcani} {et~al.} (2010) to probe to lower SFRs, but here we
show that the difference between the field and cluster galaxies is detectable even among the most 
actively star-forming galaxies. In addition, the differences
that we measure are detected with higher significance according to a K-S test ($>3\sigma$ versus $2\sigma$).  
This is likely due to the larger sample size used in this study which we obtain by including
the photometric cluster members as well as the spectroscopic cluster members.

\subsection{Spatial Distribution of IR Galaxies}
\label{radial}

\begin{figure*}
\plottwo{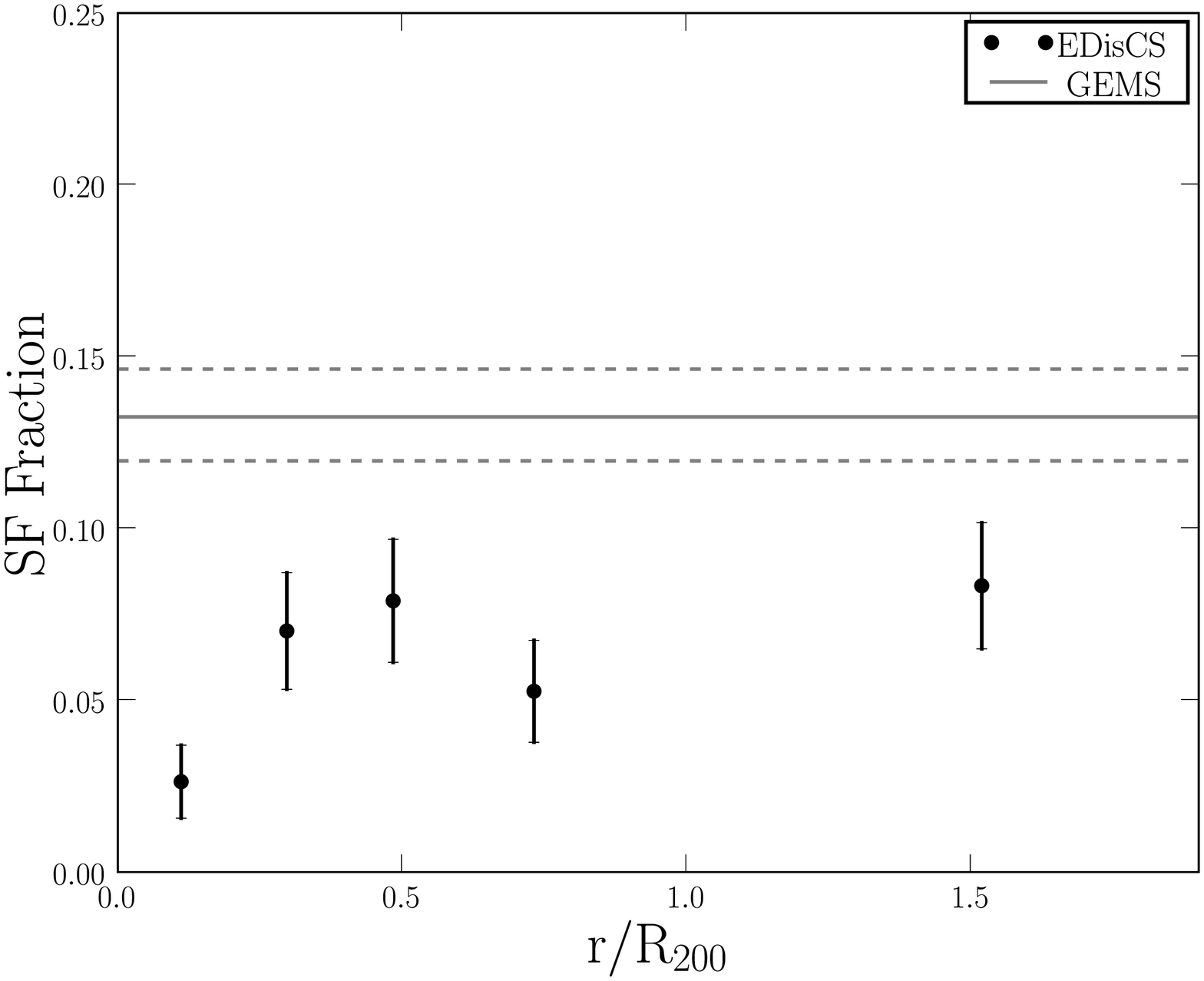}{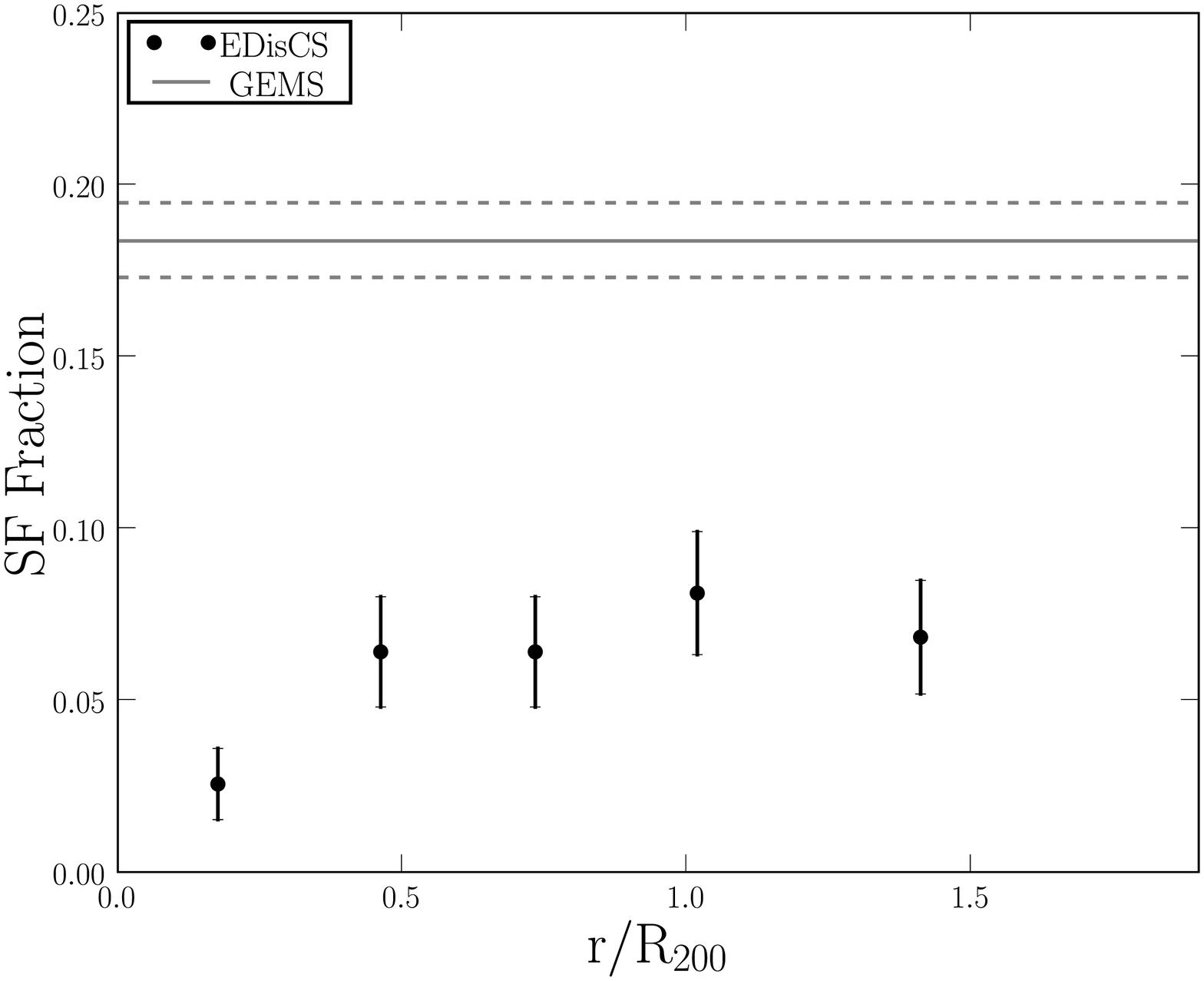}
\caption{
(Left) Fraction of IR galaxies ($\rm log_{10}(L_{IR}/L_\odot) > 10.75$, 
$M_V < -19$) in equally-populated bins as a function of projected separation from the 
cluster center for $z < 0.6$ clusters only.  
Errorbars show Poisson errors.  The gray line shows the fraction of IR galaxies
in the $0.42 < z < 0.6$ GEMS field sample, and 
the dashed lines show the $1-\sigma$ Poisson errors.  
The fraction of IR galaxies in the $z < 0.6$ clusters 
increases with projected cluster-centric radius out to 
0.5$\times$\rtwo \ and then levels.  
(Right) Fraction of IR galaxies 
($\rm log_{10}(L_{IR}/L_\odot) > 10.95$ and $M_V < -19$)
in equally-populated bins as a function of projected separation from the 
cluster center for $z > 0.6$ clusters only.  
The gray line shows the fraction of IR galaxies
in the $0.6 < z < 0.8$ GEMS field sample, and 
the dashed lines show the $1-\sigma$ Poisson errors.  
The fraction of IR galaxies increases with projected cluster-centric radius out to 
0.5$\times$\rtwo \ and then levels.  
\label{sffdrlz}}
\end{figure*}

To probe the spatial distribution of the IR galaxies, 
we calculate the fraction of 24\micron-emitting members 
as a function of projected radius from the cluster
center.  We again split the sample at $z = 0.6$, and we use 
a lower \lir \ cut of $\rm log_{10}(L_{IR}) > 10.75$ for the $z < 0.6$ 
sample to improve statistics.  We apply the same magnitude and 
\lir \ cuts to the $z < 0.6$ and $z > 0.6$ GEMS galaxies.  So while the
results for the two epochs are not directly comparable, the cluster and 
field fractions in each epoch are.  

We show the results for the $z < 0.6$ and $z > 0.6$ samples in the left
and right panels of Figure \ref{sffdrlz}, respectively.
In both the lower and higher-$z$ panels, 
the fraction of IR cluster galaxies 
increases with projected radius and remains systematically lower 
than the field fraction over the area probed.  
The offset between the cluster and field is greater for the higher-$z$
clusters, but this may be due to the higher \lir \ cut that is applied.
The results are consistent with many other cluster studies (e.g., {Balogh} {et~al.} 1997; {Lewis} {et~al.} 2002; {G{\' o}mez} {et~al.} 2003; {Rines} {et~al.} 2005).

\subsection{Redshift Evolution of IR Activity}
 \label{zevol}
\begin{figure*}[h]
\plotone{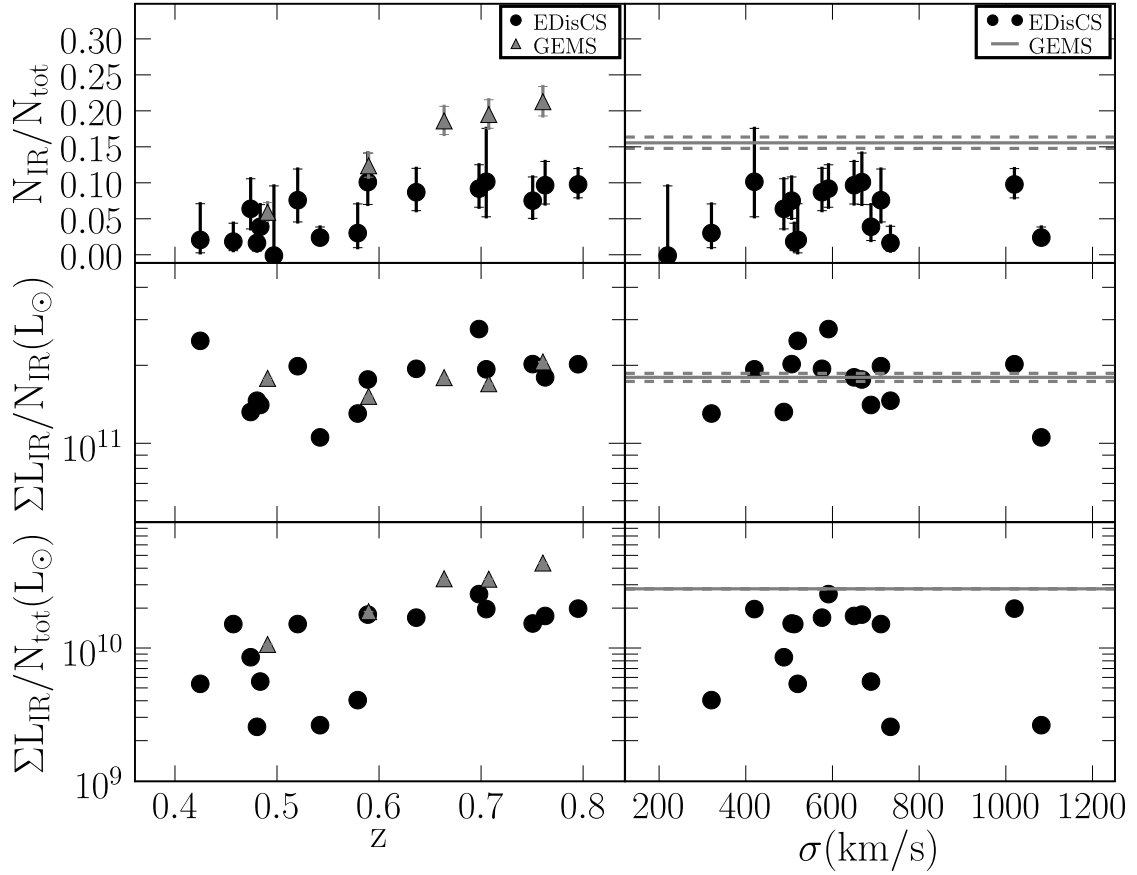}
\caption{Fraction of IR galaxies ($\rm log_{10}(L_{IR}/L_\odot) > 10.95$, $M_V < -19$, 
$\rm dr < R_{200}$) 
versus cluster redshift (top left) and cluster
velocity dispersion (top right).  Total \lir \ divided by the total number of 
IR-emitting members versus 
cluster redshift (middle left) and cluster velocity dispersion (middle right).
Total \lir \ divided by the total number of members versus 
cluster redshift (bottom left) and cluster velocity dispersion (bottom right).  
In all panels, the filled
circles show quantities for the EDisCS clusters.
The gray triangles (left panels) and
gray lines (right panels) show the 
values for the GEMS field galaxies.  Star-formation activity is correlated
with cluster redshift but not with cluster mass, and the star-formation activity of cluster galaxies 
is systematically lower than field galaxies.  Cluster-to-cluster variations are large, illustrating the need
for large samples.}
\label{fracIRevol}
\end{figure*}

To examine trends in the star-forming population 
with cluster redshift and mass, 
we characterize the star-forming activity of the clusters in 
three ways:  
 the fraction of 24\micron-emitting galaxies, 
the average \lir \ of the 24\micron \ galaxies, and the total 
\lir \ averaged over all cluster members.  
These quantities are shown in the top, middle, and bottom panels of 
Figure \ref{fracIRevol}, respectively, versus cluster redshift (left) and velocity
dispersion (right).  
We calculate the number of IR galaxies and the total IR luminosity per cluster
using only those galaxies with $M_V < -19$ and 
$\rm log_{10}(L_{IR}/L_\odot) > 10.95$, 
and we normalize by the total number of 
cluster galaxies with $M_V < -19$ ($\rm N_{tot}$).
Note that we also calculate the various measures of star-formation activity using an 
evolving $M_V$ limit as done in
{Poggianti} {et~al.} (2008).  Specifically, we vary the $M_V$ cut 
from $-20.5$ at $z=0.8$ to $-20.1$ at $z=0.4$ to account for passive evolution.  
The results are not significantly impacted by the evolving magnitude cut, so 
we adopt a constant magnitude limit of $M_V < -19$.

We limit the analysis to galaxies that have a projected cluster-centric radius less than \rtwo, 
where \rtwo \ is the radius inside which the enclosed density is 200 times the critical
density and approximates the virial radius of the cluster.  
The relationship between \rtwo \ and velocity dispersion is shown
in Equation \ref{rtwoeqn} below\footnote{This derivation of \rtwo \ 
assumes the line-of-sight velocity dispersion
is related to the circular velocity by $\sigma_x = v_c/\sqrt{3}$ rather than $\sigma_x = v_c/\sqrt{2}$ as one would expect if
galaxies are orbiting isotropically in a single isothermal sphere ({Finn} {et~al.} 2008).  Nevertheless, we adopt this 
$\sqrt{3}$ scaling between $\sigma_x$ and $v_c$ 
for consistency with previous \edi \ publications and note that our value of \rtwo \ is roughly
20\% higher as a result.}, and we refer the reader to 
{Finn} {et~al.} (2005) for a complete derivation of this relationship.
\begin{equation}
R_{200} = 2.47 \ \frac{\sigma_x}{1000~{\rm km/s}} \ \frac{1}{
\sqrt{\Omega_\Lambda + \Omega_0 (1+z)^3}}\  h_{70}^{-1} \ {\rm Mpc}.
\label{rtwoeqn}
\end{equation}

We show the three measures of star-formation activity in 
Figure \ref{fracIRevol}.  
The corresponding quantities for the GEMS galaxies (gray
triangles and lines) are calculated using the same 
magnitude and \lir \ cuts that were applied to the 
cluster galaxies.  
One main result from this figure is that 
the fraction of IR cluster galaxies lies significantly 
below that of the field, consistent with previous studies (e.g., {Tran} {et~al.} 2009);
the overall fraction of IR emitting galaxies is 
$6 \pm 1$\% for the cluster sample and 
$14 \pm 1$\% for the $0.42 < z < 0.8$ GEMS sample.
Furthermore,
all three measures of star-formation activity show that the higher-redshift 
clusters
have a higher level of star-formation activity than the lower-redshift 
clusters.  
The fraction of IR galaxies shows the most 
significant correlation ($\sim 3\sigma$) versus redshift as indicated by a Spearman rank test. 
The right panels of
Figure \ref{fracIRevol} show that LIRG activity is not correlated with cluster
velocity dispersion.  Note that we are probing only the most actively 
star-forming galaxies; {Poggianti} {et~al.} (2006) find that star-formation properties
of less active galaxies do depend on cluster mass.

If we split the cluster sample into two redshift bins, we are able to
detect redshift-evolution with higher statistical significance.  
For example, the average fraction
of LIRGs within \rtwo \ is $9^{+1}_{-1}$\% for the $z > 0.6$ clusters and 
$4^{+1}_{-1}$\% for the $z < 0.6$ clusters.  
If we split the field sample at $z = 0.6$, we find that
the fraction of GEMS LIRGs drops from $18 \pm 1$ to $8 \pm 1$\%.  
The conclusion then is that the higher-redshift galaxies in all environments
have a higher fraction of LIRGs and the amount of obscured star formation declines 
significantly over the 2.4 Gyr timeline spanned by the \edi \  clusters.
In addition, the IR fraction declines by
a factor of 2.25$^{+1.08}_{-0.65}$ in the clusters and 2.25$^{+0.46}_{-0.36}$ 
in the field.  The cluster and field decline rates agree within the errors, and thus 
the rate at which the LIRG population declines is not affected 
by the cluster environment.  

\begin{figure}[h]
\plotone{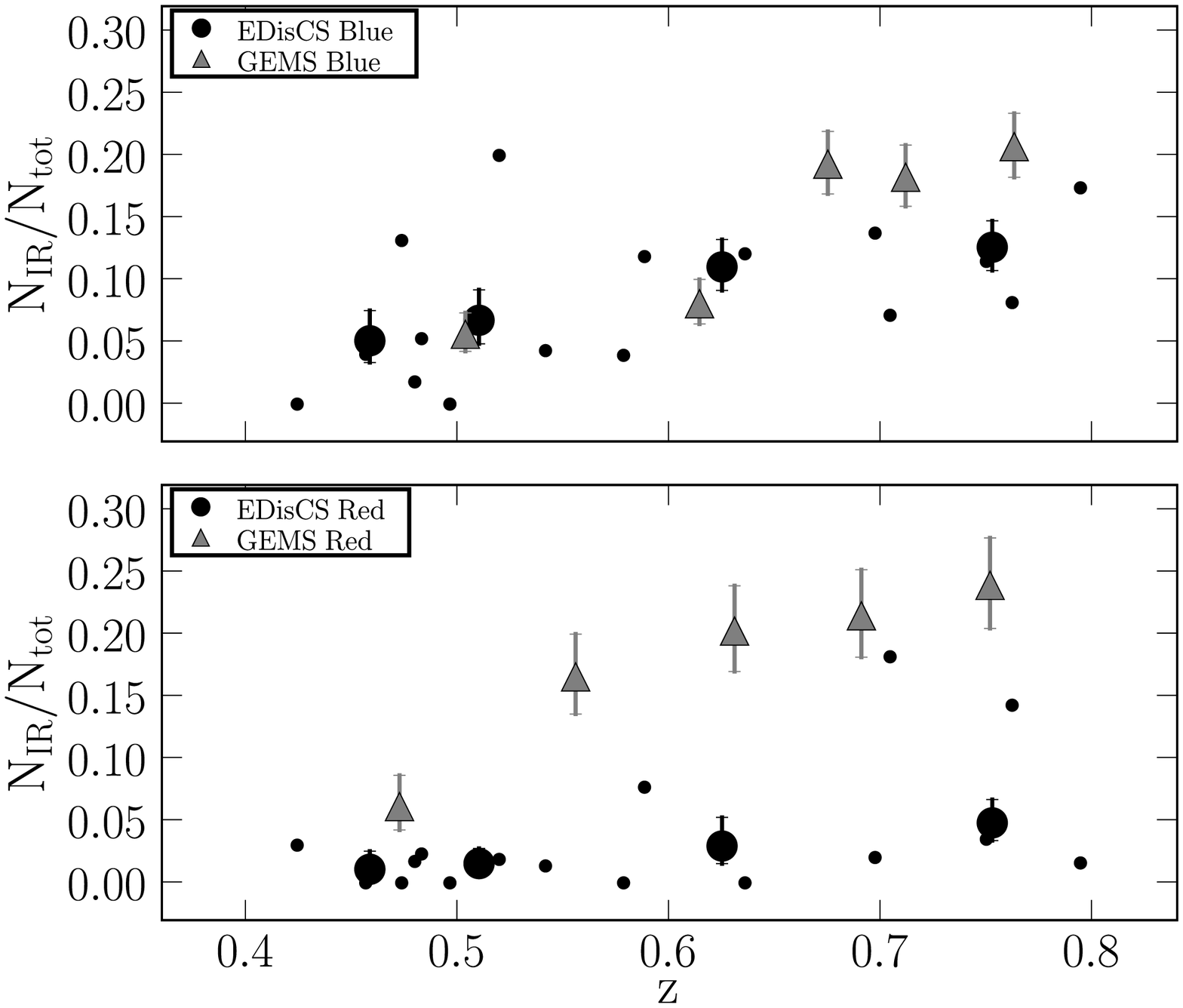}
\caption{
Fraction of IR galaxies ($\rm log_{10}(L_{IR}/L_\odot) > 10.95$, $M_V < -19$, 
$\rm dr < R_{200}$) 
versus cluster redshift considering only blue galaxies (top) and 
red galaxies (bottom).
The small circles show values for individual clusters, where errorbars are omitted for clarity.
The large circles show binned fractions for clusters.
The gray triangles show the 
values for the GEMS field galaxies.  
\label{sffcolor}
}
\end{figure}

In Figure \ref{sffcolor}, we compare the redshift-evolution of the LIRG fraction 
of the blue and red galaxies separately.  To calculate the
fractions, we divide the number of blue LIRGs by the total number of blue galaxies, and
likewise for the red galaxies.
The results for the blue galaxies are shown in the top panel.  
While a number of individual clusters (small circles) show a blue LIRG fraction
that exceeds the field value, when the data are binned to improve statistics, the
resulting LIRG fractions (large circles) among the blue cluster galaxies are 
consistent with the field values.  
The large cluster-to-cluster variations illustrate 
the need for large samples of clusters to accurately characterize  
star-formation activity in cluster environments; observations of only one cluster might
erroneously lead the observer to infer that clusters contain an excess of IR 
galaxies relative to the field.
The evolution of the red LIRG fraction is shown in the bottom panel, and
the red cluster galaxies lies systematically below that of the
field.  The difference between the cluster and field red LIRG fractions is large.  
The low fraction of red cluster LIRGs is due in part to the higher overall
fraction of red galaxies in clusters, meaning the number of red LIRGs is normalized
by a larger number of red galaxies.  However, this does not fully account for the
difference, and we will further investigate the differences in the red cluster/field
populations in a future paper.


\begin{figure}[h]
\plotone{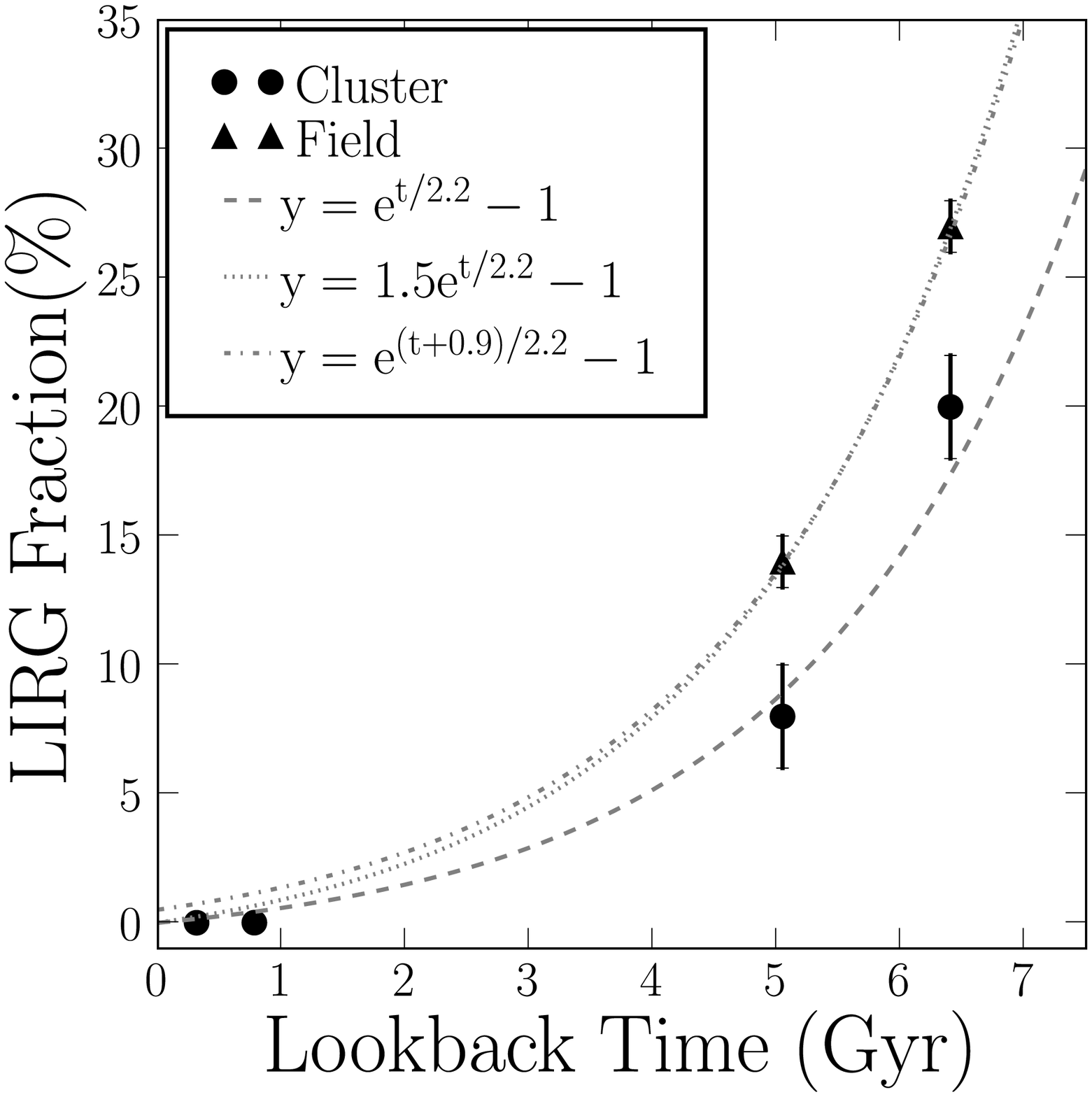}
\caption{Fraction of IR galaxies (\mvcut,  $\rm log_{10}(L_{IR}/L_\odot) > 10.95$, 
dr$<$\rtwo) versus lookback time for clusters (circles) and field (triangles) galaxies.
The dashed line shows an exponential with an e-folding time of 2.2~Gyr, which fits the cluster
data (filled circles) reasonably well.
The dot-dashed line shows the same exponential decline but delayed by 0.9~Gyr, while the dotted
line shows an exponetial with a faster decline rate.  A low-redshift baseline is
needed to better constrain the evolution of IR galaxies in the field.  
\label{sfrdecline}}
\end{figure}

$Spitzer$ studies of local clusters reveal very few galaxies with 
$\rm log_{10}(L_{IR}/L_\odot) > 10.95$ ({Bai} {et~al.} 2006, 2009).  In fact, 
the fraction of such galaxies with $M_V < -20.1$ \  within \rtwo \ is
0/274 for Coma and 1/288 for Abell 3266 (L. Bai 2009, private communication).
While Coma and Abell 3266 are more massive than the \edi \ clusters and thus might not be a fair
baseline because of this mass mismatch, we find no dependence
of the LIRG fraction on cluster velocity dispersion within the \edi \ sample, and so we proceed
with the comparison.  In Figure \ref{sfrdecline} we show the fraction of LIRGs versus
lookback time for the \edi \ and local clusters.  The dashed line shows an
exponential with an e-folding time of 2.2~Gyr.  While this is by no means a unique
fit to the data, it does show that the drop in LIRG fraction is consistent
with an exponential decline.  
In their study of eight massive $0.02 < z < 0.83$ clusters, 
{Saintonge} {et~al.} (2008) also find that the fraction of IR galaxies climbs steadily with redshift.

For comparison, we also show the fraction of IR galaxies in the GEMS sample in Figure \ref{sfrdecline}.  
As an illustration, we show two curves with the field data that depict two different evolutionary 
scenarios.  The first curve (dot-dashed line) is the same
function as for the clusters, but the decline in the field is delayed by 0.9 Gyr with respect
to the clusters.  The second curve (dashed line) shows the field LIRG population 
declining at a faster rate than the cluster LIRGs; this model produces the same behavior over the
epoch probed by the GEMS galaxies but predicts fewer LIRGs in local field environments.  
A low-redshift baseline is needed to constrain the decline rate of field LIRGs.  

Comparison with previous \edi \ studies suggests that LIRGs decline at a faster rate than normal star-forming galaxies.
In a study comparing three $z > 0.6$ \edi \ clusters to a large sample of SDSS clusters, {Finn} {et~al.} (2008)
find that the fraction of \ha-emitting galaxies declines by a factor of $6\pm3$.  
Comparison with results from low-redshift clusters ({Bai} {et~al.} 2007, 2009) suggest that
the fraction of LIRGs has decreased by a factor of $\sim$100 during the 
same time period.  
This disparity in evolution rates is likely due to the systematic decline in
star-formation activity of all star-forming galaxies.  For example,
if we assume that the IR luminosity of all galaxies is fading at the same rate, then
the IR luminosity function will shift systematically to lower 
luminosities with decreasing redshift (e.g. {Le Floc'h} {et~al.} 2005).  The most actively
star-forming galaxies - the ones we detect in the \edi \ clusters with {\it Spitzer} - 
will not be present at lower redshift.  In contrast, the number density of galaxies at 
lower IR luminosities does not change as dramatically.  
We will explore the evolution of the cluster IR luminosity function 
in more detail in a future paper.

\section{Discussion}
\label{discussion}

\subsection{The Connection between LIRGs and e(a) Galaxies}
The MORPHS collaboration first termed the spectral class e(a) to denote galaxies
that show an A-star spectrum with emission lines
({Dressler} {et~al.} 1999; {Poggianti} {et~al.} 1999).  These galaxies are often associated with 
dusty starburst galaxies in the local universe ({Poggianti} \& {Wu} 2000).  
{Poggianti} {et~al.} (2009) use the \edi \ sample to further investigate the environments of these
dusty starburst galaxies, and we compare the properties of this optically-defined sample to our
IR-selected galaxies.  Poggianti et al. find that e(a) galaxies make up 12\% of the $M_V < -20.1$ 
cluster population; we find that LIRGs make up $\sim$15\% of $M_V < -20.1$ cluster members, entirely consistent
with the assumption that e(a) galaxies are dusty starburst and are the same galaxies that
we are detecting as LIRGs. 
Furthermore, of the LIRGs that have existing optical spectroscopy, roughly three-quarters exhibit
e(a) spectra.  
Finally, e(a) galaxies are found 
least frequently in the centers of clusters.  Again, this is entirely consistent with the environmental
dependence we find for LIRGs.
{Dressler} {et~al.} (2009a) also find that $Spitzer$-detected galaxies in intermediate-redshift clusters
exhibit predominantly e(a) spectra.

{Dressler} {et~al.} (2009b) show that e(a) galaxies in a $z=0.4$ cluster have IR SFRs that are on
average 4 times greater than SFRs derived from [OII] emission.  
For our sample, the median ratio of SFRs derived 24\micron \ emission with
dust-corrected SFRs derived from [OII]  is 2.9 ({Vulcani} {et~al.} 2010).  
The large ratio of IR to optical SFRs is not surprising given the high SFRs of the
IR galaxies (e.g. {Zheng} {et~al.} 2006).

\subsection{Do Clusters Cause LIRGs?}

Some galaxy evolution models predict a cluster-induced burst of star formation 
as a galaxy falls into a cluster for the first time and 
the interstellar medium is compressed by the intra-cluster medium (e.g., {Bekki} \& {Couch} 2003).
We find no evidence that clusters contain an overabundance of LIRGs relative to the field,
at least within \rtwo.  
In contrast,  results for other $z \sim 0.8$ galaxy clusters presented by 
{Marcillac} {et~al.} (2007) and {Bai} {et~al.} (2007) conclude that LIRGs occur more frequently in clusters.  
However, these authors use the
surface density of LIRGs to quantify density without normalizing by the 
total number of cluster galaxies.  Thus, their inferred overdensity of LIRGs
relative to the field is likely due to the fact that the surface density of all galaxies
is higher in the vicinity of the cluster.  In addition, as shown in Figure \ref{fracIRevol}, 
cluster-to-cluster variations are large, and thus large samples of clusters are needed to 
accurately quantify the star-formation properties of clusters.

\subsection{Mechanisms to Explain Declining SFRs}

The SFRs of both field and cluster galaxies have declined dramatically since $z \sim 1$, and 
we have yet to address the possible mechanisms that can shut off star formation in these galaxies.
This brings us full-circle to the goals outlined in \S\ref{intro}.

Before proceeding we note that different mechanisms may be responsible for 
the star-formation and morphological transformations in clusters.  We know that
between the epoch of the \edi \ clusters and the present universe, 
the fraction of cluster spirals declines while the population of S0s increases.
{Desai} {et~al.} (2007) find no significant change in S0 fraction within \edi \ sample, but we observe a decline in SFRs.  
This implies SFRs are changing faster than
morphologies, and in fact the SF and morphology evolution might be 
driven by two different mechanisms.
This supports results of other cluster studies ({Poggianti} {et~al.} 1999, 2006; {S{\'a}nchez-Bl{\'a}zquez} {et~al.} 2009; {Geach} {et~al.} 2009).
In the remainder of this subsection, we focus solely on the processes that may be affecting
star-formation rates.

From our observations, we are able to place constraints on the
mechanism(s) responsible for the decline of luminous IR galaxies.  
We see a decline in the fraction of 
LIRGs in both the cluster and field samples, and the rate of the decline is the same
for both samples. 
Furthermore, the \lir \ distribution is the same for the field and cluster galaxies, 
but the fraction of LIRGs is lower in the clusters, 
similar to results for local clusters (e.g. {Balogh} {et~al.} 2004; {Tanaka} {et~al.} 2004; {Rines} {et~al.} 2005; {Finn} {et~al.} 2008).  
Finally, the stellar masses of the cluster LIRGs are on average higher than their 
field counterparts.
One possible mechanism that can account for the similar evolution of field and cluster
galaxies is gas depletion, where galaxies are not able to 
replenish their disk gas and have slowly declining star formation histories
from using their available gas.  The cluster environment would then have 
a lower fraction of star-forming
galaxies because the cluster galaxies are further evolved. 
Figure \ref{sfrdecline} shows one plausible scenario where the decline in field
LIRGs mirrors the decline in cluster LIRGs, 
but the field evolution is delayed by $\sim 1$Gyr.

This scenario can also explain the higher stellar masses seen in the cluster LIRGs.
For example, Fig. \ref{lirstellmassredblue} shows that SFR and stellar mass are correlated in the
sense that more massive galaxies have higher SFRs.  As gas depletion progresses, the 
SFRs of all galaxies decrease, and LIRG-levels of 
star-formation would remain only in the most massive galaxies.  
In our scenario, gas depletion is more advanced in the cluster environment, and thus one would 
also expect the host galaxies of LIRGs to be more massive.

There is evidence that
clusters affect the SFRs of some in-falling galaxies.  
For example, {Poggianti} {et~al.} (2009) and {Poggianti} {et~al.} (1999) find an excess of post-starburst galaxies
in the \edi \ clusters relative to the coeval field.  
These galaxies have had their star formation quenched abruptly within 1~Gyr prior to observation, 
and their excess relative to the field indicates 
that cluster-specific processes 
are altering the star-formation properties of some infalling galaxies.
The excess of post-starburst galaxies in clusters is small, and we don't have a large enough sample
to detect a small differential evolution between the cluster and field LIRG fraction.

{Rudnick} {et~al.} (2009) provide further evidence that clusters are actively altering infalling galaxies; 
they find that the red sequence appears to build up more rapidly in clusters than in the field.
In addition, {De Lucia} {et~al.} (2007) 
find that the rate at which the red sequence builds up depends
weakly on cluster velocity dispersion, although the interpretation of the observational
findings depends on assumptions regarding the nature of the underlying mass/luminosity
function of galaxies in various environments.

\section{Conclusions}
\label{conclusions}
We present {\it Spitzer} MIPS observations of 16 $0.4 < z < 0.8$ \edi \ clusters.
This is the first large 24\micron-survey of clusters at intermediate redshift 
and represents a significant increase in the census of star-formation
rates in dense environments.  The limits of the 24\micron \ imaging are such
that we are sensitive to only the brightest IR galaxies, and so our sample contains
mainly LIRGs rather than normal star-forming galaxies.
Our major results are summarized below.

\begin{itemize}
\item We calculate \lir \ for the clusters members and
compare to a large sample of coeval field galaxies from the literature.  While the clusters contain 
a lower fraction of IR-emitting galaxies, 
the distribution of \lir \ for the cluster galaxies is identical to that of the field galaxies.

\item The stellar masses of the \edi \ LIRGs are systematically higher than the stellar masses of the GEMS galaxies.  

\item Approximately $\sim$80\% of the 
IR galaxies live in the blue cloud and the remaining 20\% lie on the red sequence.

\item The majority of LIRGs have optical spectra that are dominated by A-stars and show some signs of
modest on-going star formation as determined by [O~II] emission 
(i.e. e(a); {Poggianti} {et~al.} 1999, 2008).  SFRs derived from IR
emission are much greater than those inferred from optical emission, with a median $\rm SFR(IR)/SFR(OII)$ 
of 2.9 ({Vulcani} {et~al.} 2010).  

\item LIRGs avoid the centers of clusters; the fraction of IR galaxies is lowest near the cluster center
($\rm dr < 0.5 \times$\rtwo) and remains below the field value at least out to $1.5\times$\rtwo.  

\item The fraction of IR galaxies 
decreases significantly over the 2.4~Gyr interval spanned by our sample, and the rate of the decline
is the same for the cluster and field populations. Comparison with IR studies of local clusters 
shows that the evolution of the cluster LIRGs is consistent with
an exponential decline with an e-folding time of 2.2~Gyr.

\item Star-formation rates are declining faster than morphologies are transforming, consistent with numerous
previous studies.

\item The similar decline of field and cluster LIRGs suggests that the mechanism driving the global decline
of SFRs is the same in the cluster and field environments.  We find gas depletion 
to be the most likely candidate, where the decline in the field is delayed by $\sim1$~Gyr with respect
to the clusters.

\end{itemize}

\acknowledgements
RAF thanks L. Bai for useful discussions and for aiding with the comparison to her work.  
BMP acknowledge financial support from ASI contract I/016/07/0.  
We thank the anonymous referee for suggestions that significantly improved the paper.
This work is based on observations made with the Spitzer Space Telescope, which is operated by 
the Jet Propulsion Laboratory, California Institute of Technology under a contract with NASA. 
Support for this work was provided by NASA through an award issued by JPL/Caltech.
The Dark Cosmology Centre is funded by the Danish National Research Foundation.




\begin{deluxetable}{lccccccc} 
\tablecaption{Summary of 24\micron \ Detections \label{detect}}
\tablehead{  \colhead{Cluster} &\colhead{z} &\colhead{$\sigma$} & \colhead{$\rm f_{min}^a$} & \colhead{$\rm f_{max}^b$}&\colhead{$\rm f_{80}^c$} & \colhead{$\rm L_{IR}(80)^d$} & \colhead{SFR$_{80}^e$}\\ &  & (km/s) & ($\mu$Jy) & ($\mu$Jy) & ($\mu$Jy) & log$\rm _{10}(L_{IR}/L_\odot)$& (\smy)} 
\startdata 
CL1216.8$-$1201 & 0.7943 & 1018$^{+73}_{-77}$ &  13  & 2347  &  75$\pm$5 & 10.91 & 13.9 \\ 
CL1354.2$-$1230 & 0.7620 & 648$^{+105}_{-110}$ &   3  & 2856  &  82$\pm$3 & 10.91 & 13.9 \\ 
CL1054.7$-$1245 & 0.7498 & 504$^{+113}_{-65}$ &   9  & 2599  &  75$\pm$3 & 10.85 & 12.3 \\ 
CL1040.7$-$1155 & 0.7043 & 418$^{+55}_{-46}$ &  39  & 1296  &  80$\pm$3 & 10.83 & 11.6 \\ 
CL1054.4$-$1146 & 0.6972 & 589$^{+78}_{-70}$ &  36  & 1068  &  72$\pm$3 & 10.78 & 10.3 \\ 
CL1227.9$-$1138 & 0.6357 & 574$^{+72}_{-75}$ &   6  & 1517  &  80$\pm$3 & 10.75 &  9.6 \\ 
CL1353.0$-$1137 & 0.5882 & 666$^{+136}_{-139}$ &  50  & 4072  & 100$\pm$3 & 10.75 &  9.7 \\ 
CL1037.9$-$1243 & 0.5783 & 319$^{+53}_{-52}$ &  40  & 1556  &  82$\pm$3 & 10.65 &  7.7 \\ 
CL1232.5$-$1250 & 0.5414 & 1080$^{+119}_{-89}$ &  43  & 1541  &  97$\pm$3 & 10.64 &  7.5 \\ 
CL1411.1$-$1148 & 0.5195 & 710$^{+125}_{-133}$ &  49  & 3694  &  97$\pm$3 & 10.59 &  6.6 \\ 
CL1420.3$-$1236 & 0.4962 & 218$^{+43}_{-50}$ &  44  & 1612  &  95$\pm$5 & 10.52 &  5.7 \\ 
CL1301.7$-$1139 & 0.4828 & 687$^{+81}_{-86}$ &  42  & 1181  &  85$\pm$3 & 10.43 &  4.6 \\ 
CL1138.2$-$1133 & 0.4796 & 732$^{+72}_{-76}$ &  43  & 1441  &  85$\pm$3 & 10.42 &  4.6 \\ 
CL1018.8$-$1211 & 0.4734 & 486$^{+59}_{-63}$ &  37  & 5146  &  80$\pm$3 & 10.38 &  4.1 \\ 
CL1059.2$-$1253 & 0.4564 & 510$^{+52}_{-56}$ &  41  & 5485  &  82$\pm$3 & 10.35 &  3.9 \\ 
CL1202.7$-$1224 & 0.4240 & 518$^{+92}_{-104}$ &  22  & 5196  &  97$\pm$3 & 10.33 &  3.6 \\ 
\enddata 
\tablenotetext{a}{Minimum 24\micron \ flux detected from sources with SNR$>$2.5} 
\tablenotetext{b}{Maximum 24\micron \ flux detected from sources with SNR$>$2.5.}
\tablenotetext{c}{24\micron \ flux corresponding to 80\% completeness limit.}
\tablenotetext{d}{L$\rm_{IR}$ corresponding to 80\% completeness limit.  Relative error is the same as for f$_{80}$.}
\tablenotetext{e}{SFR corresponding to 80\% completeness limit.  Relative error is the same as for f$_{80}$.} 
\end{deluxetable}

\end{document}